\documentclass[document,showpacs]{revtex4}
\pdfoutput=1
\usepackage{amssymb,latexsym}
\usepackage{amsmath,amsbsy,bbm}
\usepackage{ifpdf}
\usepackage{epsfig,bm}
\usepackage{graphicx,comment}
\usepackage{color}
\usepackage{soul}
\usepackage{mathtools}
\usepackage{comment}
\usepackage[normalem]{ulem}
\begin{document}

\title{Berezinskii--Kosterlitz--Thouless transition -- a universal neural network study with benchmarking}
\author{Y.-H. Tseng}
\affiliation{Department of Physics, National Taiwan Normal University,
88, Sec.4, Ting-Chou Rd., Taipei 116, Taiwan}
\author{F.-J. Jiang}
\email[]{fjjiang@ntnu.edu.tw}
\affiliation{Department of Physics, National Taiwan Normal University,
88, Sec.4, Ting-Chou Rd., Taipei 116, Taiwan}

\begin{abstract}
	Using a supervised neural network (NN) trained once on a one-dimensional lattice of 200 sites,
        we calculate the Berezinskii--Kosterlitz--Thouless phase transitions of the two-dimensional (2D) classical $XY$ and the
        2D generalized classical $XY$ models.
  In particular, both the bulk quantities Binder ratios and the spin states of
  the studied systems are employed to construct the needed configurations for 
	the NN prediction.
  By applying semiempirical finite-size scaling to the relevant data, the critical points obtained by the NN
  approach agree well
  with the known results established in the literature. This implies that for
  each of the considered models, the determination of
  its various phases requires only a little information. The outcomes presented
  here demonstrate convincingly that the employed universal NN is not only valid for the symmetry breaking related phase transitions, 
	but also
  works for calculating the critical points of the phase
  transitions associated with topology. The efficiency of the used NN in 
	the computation is examined by carrying out several detailed benchmark calculations.

\end{abstract}


\maketitle

\section{Introduction}
\label{sec:intro}

Recently, techniques of Machine Learning (ML), such as supervised neural
network (NN), autoencoder, as well as generative adversarial network (GAN)
provide alternative approaches
for studying many-body physics. In particular, these ML methods are shown
to be efficient in classifying various phases of physical systems \cite{Wan16,Oht16,Car16,Tro16,Chn16,Tan16,Nie16,Den17,Zha17,Hu17,Li18,Zha18,Bea18,Lia19,Rod19,Car19,Zha19,Don19,Tan20.1,Tan20.2,Sin20,Sch20,Carrasquilla:2020mas,Tomita:2020ylz}.
They are employed in the studies of high energy physics and astronomy as well \cite{Baldi:2014pta,Hoyle:2015yha,Mott:2017xdb,Pang:2016vdc,Sha18,Cavaglia:2018xjq,Larkoski:2017jix,Amacker:2020bmn,Aad:2020cws,Cabero:2020eik,Nicoli:2020njz}.
To conclude, ML has gone far beyond its original fields of applications
such as computer, information, and medical sciences.

The employment of NN in studying many-body systems usually consists of
three stages, namely the training, the validation, and the testing
(prediction) stages \cite{Car19,Carrasquilla:2020mas}. Here we will briefly describe
the stages of training and testing.
The standard training set typically used in a NN calculation associated
with investigating phase transition consists of several to
a few thousand real configurations \cite{Car16}. As a result,
the calculations are quite demanding. Apart from this, the NNs commonly employed
have complicated infrastructure such as the
convolutional neural network (CNN). Moreover, for any given model,
special structure of the
convolutional layer is designed for that model \cite{Car19,Carrasquilla:2020mas}. The idea of using CNN is to capture the features of
the system so that the NN can efficiently learn to distinguish various
phases. From this point of view, a CNN built for a model may not be applicable
for another system. Besides, for every studied model and each
of the considered system sizes, a separate training is required. 
Finally, for the testing stage, real configurations are employed for the prediction.
This would use huge amount of storage space if system of larger sizes
are considered.
Although the ideas from NN are shown to be able to study
the critical phenomena of phase transitions, the facts that
a training should be conducted whenever a new model or a different
system size is considered, as well as the need of large storage space
may prevent NN from being practically used in real investigations.

Because of these potential hindrances, it
is desirable to construct a NN that can be used to study various phases of
many models, and during the same time is efficient in both the
computation and the storage. It is also interesting to notice that the majority
of the studies associated with using the NN method to investigate phase transitions
have focused on modifying the infrastructure of the NNs,
and seldom provide benchmarks of the performance of the used NNs.

In Ref.~\cite{Tan21}, a simple NN trained
on a one-dimensional (1D) lattice using only two configurations as the training set is built (using Keras and Tensorflow \cite{kera,tens}).
The obtained (only one) NN successfully calculates the critical points of
several three-dimensional models. In particular, both the (spin) configurations
and the bulk observables are employed to construct the needed configurations,
which are 1D lattices of 120 sites, for the NN prediction. Since only 1D configurations
are required, the storage needed for the prediction stage is only at the
permille level of that necessary for a standard NN calculation.
An estimated benchmark also indicates the 1D NN is extremely
efficient in the training stage \cite{Tan21}. Particularly, for the 3D classical $O(3)$
model, the time needed to train a NN using the standard approach with total
1000 configurations obtained at 4 temperatures and system size $L=48$ is at least 400 times
as much as that required for the 1D NN.
It should also be pointed out that the 1D NN of Ref.~\cite{Tan21} is one
of the few supervised NN approaches for studying phase transitions
that the locations of the critical points (in the associated parameter spaces)
are not required in advance.

The phase transitions corresponding to the models considered in Ref.~\cite{Tan21}
are induced by symmetry breaking and restoration. Hence it will be
interesting to examine whether the constructed 1D NN is applicable
to locate the critical point(s) associated with topological
phase transitions. We would like to emphasize the fact that the majority of
NN studies related to topological phase transitions use
techniques of unsupervised NN instead of supervised NN. Consequently,
it is motivated to conduct a
supervised NN investigation for a topological phase transition, and particularly to examine
whether the obtained outcomes are comparable in precision with those
determined by unsupervised NN methods.
Because of the reasons mentioned above, here we have used the 1D NN built in Ref.~\cite{Tan21}
to study the phase transitions of the two-dimensional (2D) classical $XY$
and the 2D generalized classical $XY$ models.

To calculate the critical point associated with the
Berezinskii--Kosterlitz--Thouless (BKT) phase transition of the 2D
classical $XY$ model,
the standard approach is to determine the helicity modulus at finite
lattices, and then applies the expected finite-size scaling (Which contains
a logarithmic correction) to obtain the bulk critical point.
Other quantities such as Binder ratios are seldom considered for
such investigations. In this study observables which satisfy certain conditions
(detailed later) will be used. In particular, the Binder ratios will be
considered in the NN calculations. 

Remarkably, using the 1D NN as well as the configurations
constructed from the Binder ratios (and the spins),
the critical points calculated by applying semiempirical finite-size scaling
to the obtained relevant quantities are in good agreement with the known results in the literature.
Our investigations indicate that in addition to the helicity modulus, the observables
Binder ratios in conjunction with the NN techniques provides an efficient
alternative approach for studying the phase transitions of BKT type.
Although BKT type transitions are typically characterized by vortices, as we will show later, our
outcomes imply that even few percent information of the whole
spin configurations can be employed to detect the existence of a BKT transition.

The results presented here demonstrate that the used 1D NN is
valid for both symmetry breaking and topology related phase transitions.
In other words, the employed 1D NN is universal from a broaden point of view.

It is beyond doubt that the storage space needed for the prediction
in this study is much less than the methods of using real configurations
for the prediction. Apart from this, we also conduct some benchmark
calculations to examine the efficient of our methods in both
the training and the prediction stages.
The results imply potentially a factor of at least several hundred to few
thousand in efficiency is gained for our methods.

\section{The constructed supervised Neural Network}

It should be pointed out that we hope to construct a supervised NN that is
applicable to as many models as possible when the phase transitions are concerned.
During the same time, the efficiency in both the computation and the
storage space should not be sacrificed. Based on such thoughts, in Ref.~\cite{Tan21}
a NN trained on a 1D lattice of a given fixed size is built. Moreover,
it is shown as well that such a NN can be employed to determine
the critical points of several 3D systems successfully. Therefore,
in this study, the 1D NN of \cite{Tan21} is recycled and is adopted
here. In other words, no NN, except those for the benchmarking calculations, is trained in this investigation.

\begin{figure*}
	\begin{center}
		\includegraphics[width=0.95\textwidth]{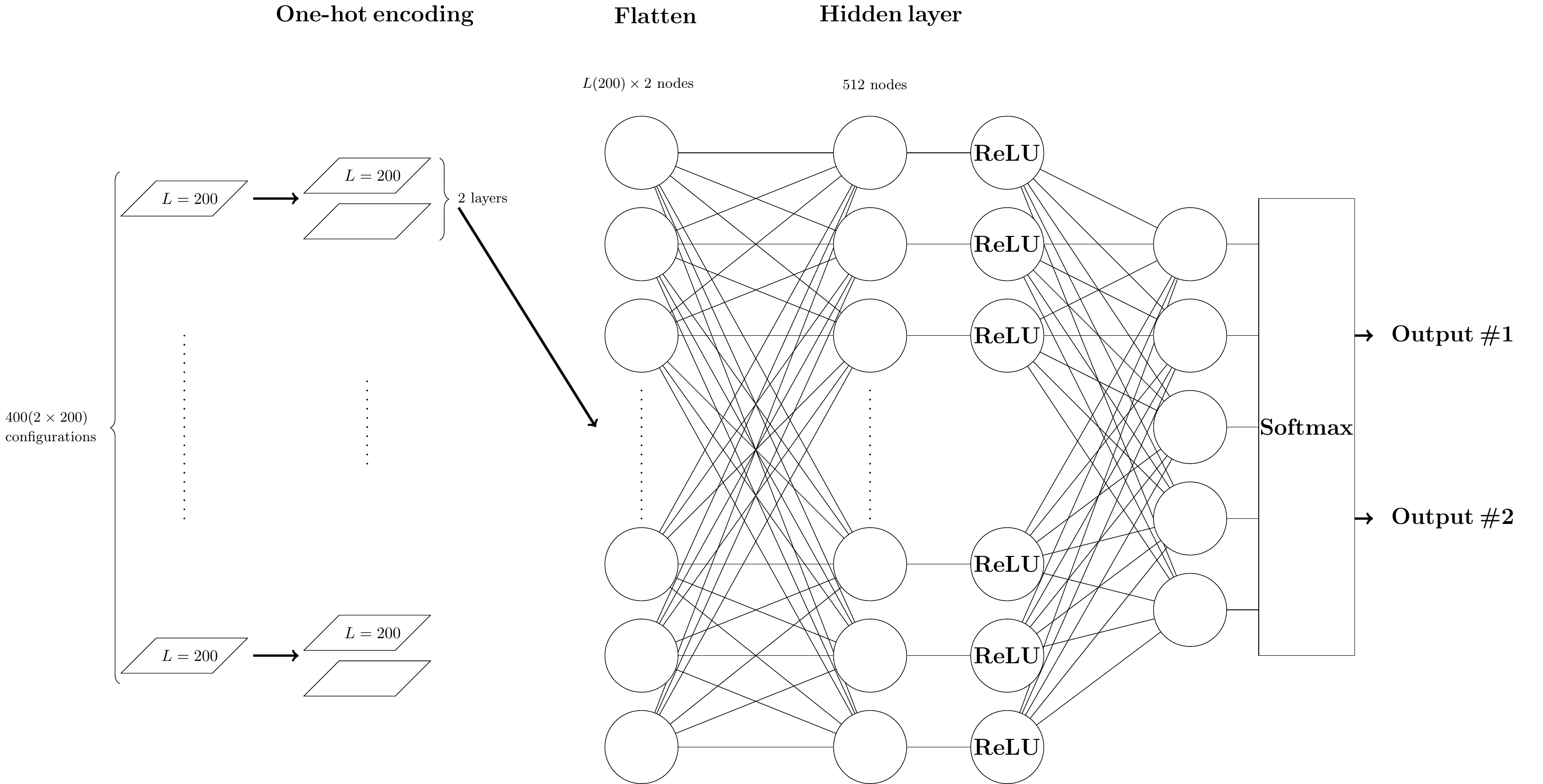}
	\end{center}\vskip-0.7cm
	\caption{The NN (MLP), which consists of one input layer, one hidden layer,
		and one output layer, used here and in Refs.~\cite{Tan20.1,Tan20.2,Tan21}.
		The training set are
		made up of 200 copies of only two configurations.
		There are 512 nodes in the hidden layer and each of these
		nodes is independently connected to every object in the training set.
                The steps of one-hot encoding and flatten are applied as well. The activation
		functions ReLU and softmax are employed in the associated layers.
                Finally, the output layer consists of two-component vector(s).
		The figure is reproduced from Refs.~\cite{Tan20.1,Tan20.2,Tan21}}  
	\label{MLP}
\end{figure*}

To make current paper self-contained, some technical details including
the constructed 1D NN are briefly quoted from Ref.~\cite{Tan21}.
The 1D NN employed here is a multilayer perceptron (MLP) which consists
of only one input layer, one hidden layer of 512 
independent nodes, and one output layer. Moreover, the algorithm considered
for the whole NN steps is the minibatch. The adam and the categorical cross entropy
are the optimizer and the loss function used.
$L_2$ regularization is applied so that the overfitting can be prevented.
The activation functions considered between various layers are the ReLU and the
softmax functions. The details of the constructed MLP
are shown in Fig.~\ref{MLP}. Notice Fig.~\ref{MLP} indicates that
the techniques of one-hot encoding and flatten are also included in the
infrastructure of the used NN. Refs.~\cite{Tan20.1,Tan20.2,Tan21} contain 
more detailed descriptions of the employed NN .

For completeness, the training and the prediction strategies
used in Ref.~\cite{Tan21} will be introduced (quoted) for the benefit of readers.
For the training set, instead of using
real configurations obtained from the simulations,
two artificial made configurations are employed.
Particularly, one of the configuration has 0 as the value
for each of its site, and every spot of the other configuration
takes the number of 1. Because of the features of the considered training set,
the corresponding labels are $(1,0)$ and $(0,1)$. 

We would like to emphasize the fact that the employed NN was trained
on a 1D lattice of 200 sites. Hence the training process
takes much less computing resources than (any) other known schemes in the
literature. We will provide some benchmarks later to demonstrate this fact.
The considered approach for the NN prediction will be detailed
later as well in the relevant sections.

It should be pointed out that 10 sets of random seeds are used so that there will be
10 MLPs. Each MLP will lead to an outcome which is the average over few hundred to
several thousand results. The (final) main results quoted here
are based on the means and the mean errors of the 10 outcomes. In other words,
the errors shown here are associated with the NN, not the raw data.

\section{The microscopic models and observables}

The Hamiltonian $H$ of the studied 2D generalized classical $XY $model on the square lattice
is given by \cite{Has05,Can14,Can16}
\begin{equation}
H = - \sum_{\left< ij\right>} \Delta\cos\left(\theta_i - \theta_j\right) - \left(1-\Delta\right)\cos\left(q\theta_i-q\theta_j\right),
\label{eqn1}
\end{equation}
where the summation is over nearest neighbor $i$ and $j$, $q$ is some
(positive) integer, and $0\le \theta_i \le 2\pi$. When $\Delta=1$, the above
Hamiltonian
reduces to that of the standard 2D classical $XY$ model. In our study, we will consider the case of $\Delta = 0.25$
and $q=3$.

Relevant observables used in this study are the first and the second Binder
ratios which are given as
\begin{eqnarray}
Q_{1,j} = \langle |m_j| \rangle^2/ \langle m_j^2\rangle \,\,\text{and}\,\, Q_{2,j} = \langle m_j^2\rangle^2 / \langle m_j^4 \rangle,
\end{eqnarray}
respectively. Here $j=1$ or (and) 3, $m_j = \frac{1}{L^2}\sum_k \exp\left(i\theta_k j \right)$
and $L$ is the linear box size of the system \cite{Bin81}.
Unless confusion may arise, the subscript $j$ in $Q_{1,j}$ and $Q_{2,j}$ will be
omitted.

\section{Benchmarks of the performance of the employed 1D NN}

Before presenting the numerical results, we would like to show
some benchmarks of the performance of the employed 1D NN here.
The considered model for the benchmark investigation is the 2D classical XY model.
We have carried out the associated calculations using both a CNN and a MLP.
Comparison between the performance regarding the training of the employed 1D NN and a autoencoder is shown as well.
The benchmark investigations are done on a server with 24 cores (two opteron 6344) and 96G memory. 
The installed version of tensorflow uses multiple CPUs for the calculations by default. Hence 
we keep only one training job running on the server at a time.
It should be pointed out that the outcomes of the benchmarks may change due to some facts such as what machine
is used for the calculations (CPU or GPU) and the jobs loading in that machine. Installing
the tensorflow with different (versions of) libraries and how the executed codes are implemented would affect the benchmarks as well.
Nevertheless, the benchmarking shown here
provides useful information regarding the performance of the considered NNs.
Finally, tunable parameters associated
with NN not mentioned explicitly in the text are the default ones.  

Unlike the main outcomes which will be demonstrated later, the shown mean errors in this section
are associated with the raw data, i.e. we use one NN, not 10 NNs for the calculations. 

\subsection{Benchmarks using a MLP}

The employed MLP consists of one input layer, one hidden layer of 512 nodes, and one output layer.
The epochs and the batch size used in the calculations are 800 and 40, respectively. Moreover, three calculations using
various training strategies are conducted. The details are as follows.

\begin{enumerate}
\item{For each of $L=64$, 96, and 128, 12 temperatures (6 below and 6 above the critical point) with each temperature
  having 1000 real configurations are considered. In addition, the variable $\theta_i$ at each site $i$ is converted to
  $\theta_i\,\text{mod}\,\pi$, and the resulting configurations are used as the training set. The step of one-hot encoding
  is employed as well.}
  \item{For each of  $L=64$, 96, and 128, 12 temperatures (6 below and 6 above the critical point) with each temperature
    having 1000 real configurations are used as the training set. The raw spin configurations are used directly as the training set
    and the step of one-hot encoding is not employed.
    Based on the relevant publications in the literature, this training
  procedure is the standard method of training the needed NN for studying the 2D classical $XY$ model. }
  \item{200 copies for each of the two artificial configurations described in the previous section are used as the training set.
  One-hot encoding is considered.}
\end{enumerate}

For convenience, These three investigations (from top to bottom) will be name the first, the second, and the third (training) methods.
The total time takes to complete the whole training processes for each of the three different training strategies
is as follows.

\begin{enumerate}
\item{First training method: 2233.2 minutes (The time needed for training the $L=128$ data is 1227.3 minutes).}
\item{Second training method: 1078.4 minutes (The time needed for training the $L=128$ data is 605.27 minutes).}
\item{Third training method: 1.2 minutes.}
\end{enumerate}

The results indicate that the time required to train a MLP using the first
method is about 2000 times as much as that needed to train a MLP with the third method.
The time required for the prediction associated with the first training method
is also several times as much as that needed for the prediction related to the third training approach.

\begin{figure}
	\begin{center}
		
		\vbox{
			\includegraphics[width=0.45\textwidth]{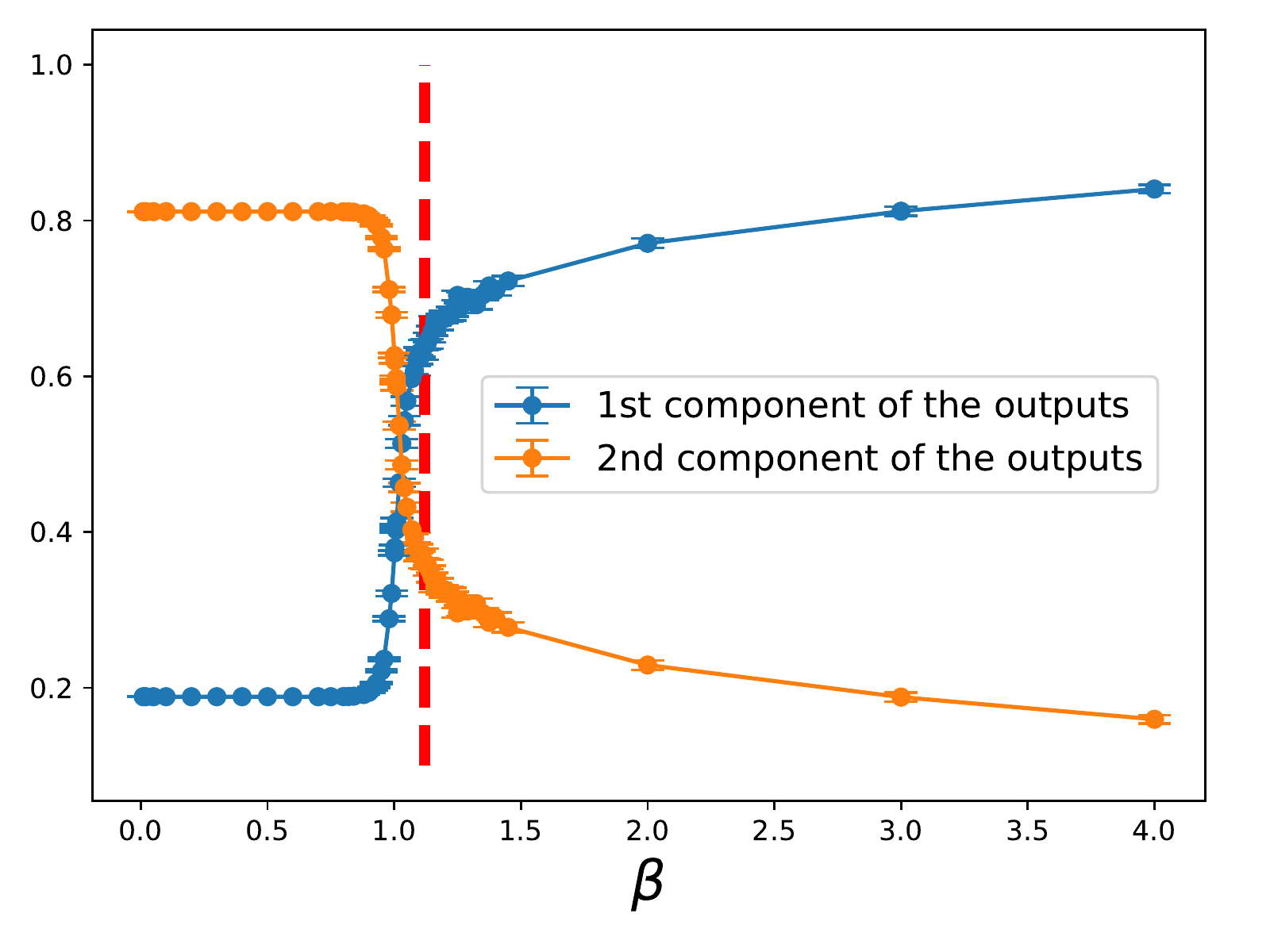}
			\includegraphics[width=0.45\textwidth]{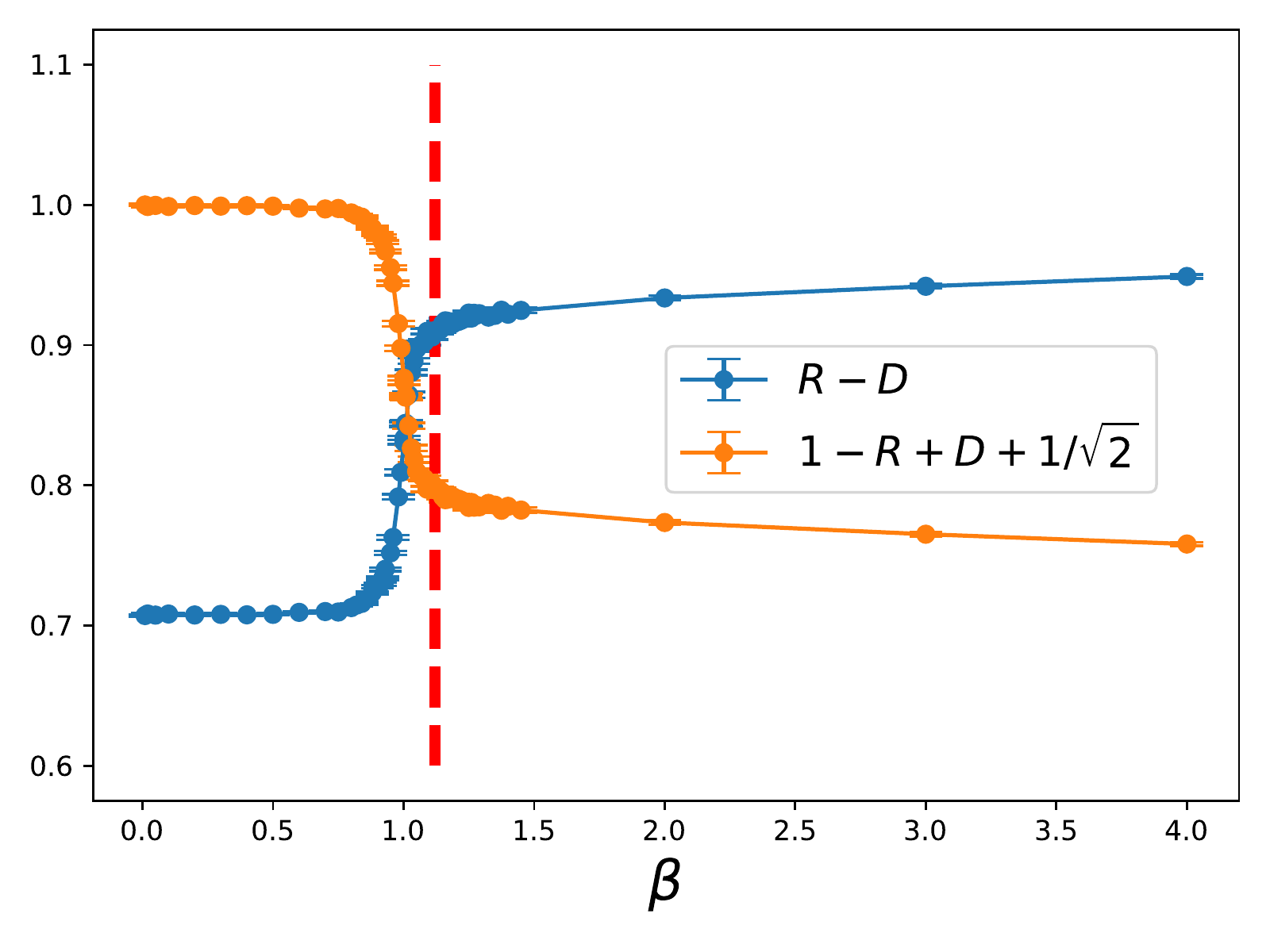}
		}
               \end{center}
	\caption{The MLP determination of the finite size critical inverse temperatures
          for the 2D classical $XY$ model (The system size is $L=128$). The left and the right panels
          correspond to the first and the third training methods, respectively, see the main text for the details.
        Here $R$ are the norms of the output vectors and $D$ is the difference between the $R$ of the smallest $\beta$ and $1/\sqrt{2}$.}
	\label{Benchmark_1}
\end{figure}

The NN prediction using these three trained NN and the real spin configurations of $L=128$
leads to the outcomes demonstrated in Fig.~\ref{Benchmark_1}. For the NN obtained
using the third training method, each of the configurations employed for the prediction is
built by firstly choosing 200 spins (randomly) from the raw $128^2$ spins, and then 
using the associated $\theta\,\text{mod}\,\pi$ to construct the needed 200 sites configuration (for the prediction).

The left and right panels of Fig.~\ref{Benchmark_1} are associated with the first and the third training methods, respectively.
In addition, the intersecting points and the vertical dashed lines in these two panels are the NN predicted and the expected
$\beta_{\text{KT}}$ for the 2D classical $XY$ model, respectively. The outcomes related to the second training method are not shown because
the training is not
successful. Indeed, for the second training approach, the simple MLP used for the training can hardly learn any information of the system
from the raw spin configurations. Finally, the fact that the use of $\theta\,\text{mod}\,\pi$ instead of $\theta$ can lead to an
outcome that a simply
built NN is capable of detecting the BKT transition is quite remarkable.

It is beyond doubt that the results in Fig.~\ref{Benchmark_1} indicate the
accuracy of the finite size critical $\beta$s
obtained by the first and the third approaches are comparable with each other.

\subsection{Benchmarks using a CNN}

Ideally, the benchmark study should be conducted with a CNN that has been used in the literature. However,
some details are typically missed in the relevant publications, for instance, the explicit expressions of the filters considered
in Ref.~\cite{Bea18} are not known. As a result, here we construct our own CNN for the investigations. The CNN considered here has
one input layer, one convolutional layer, one average pooling layer, and one output layer.

For the CNN calculations, we use three training (and prediction) procedures as those associated with the
MLP study described above. Notice the filter of the convolutional layer for
the third training method is one-dimension, and the filters related to the second and the third training approaches are two-dimension.
The resulting total time for each of the training strategies is as follows.

\begin{enumerate}
\item{First training method: 580.9 minutes.}
\item{Second training method: 518.3 minutes.}
\item{Third training method: 1.0 minutes.}
\end{enumerate}

\begin{figure}
	\begin{center}
		
		\hbox{
		  \includegraphics[width=0.315\textwidth]{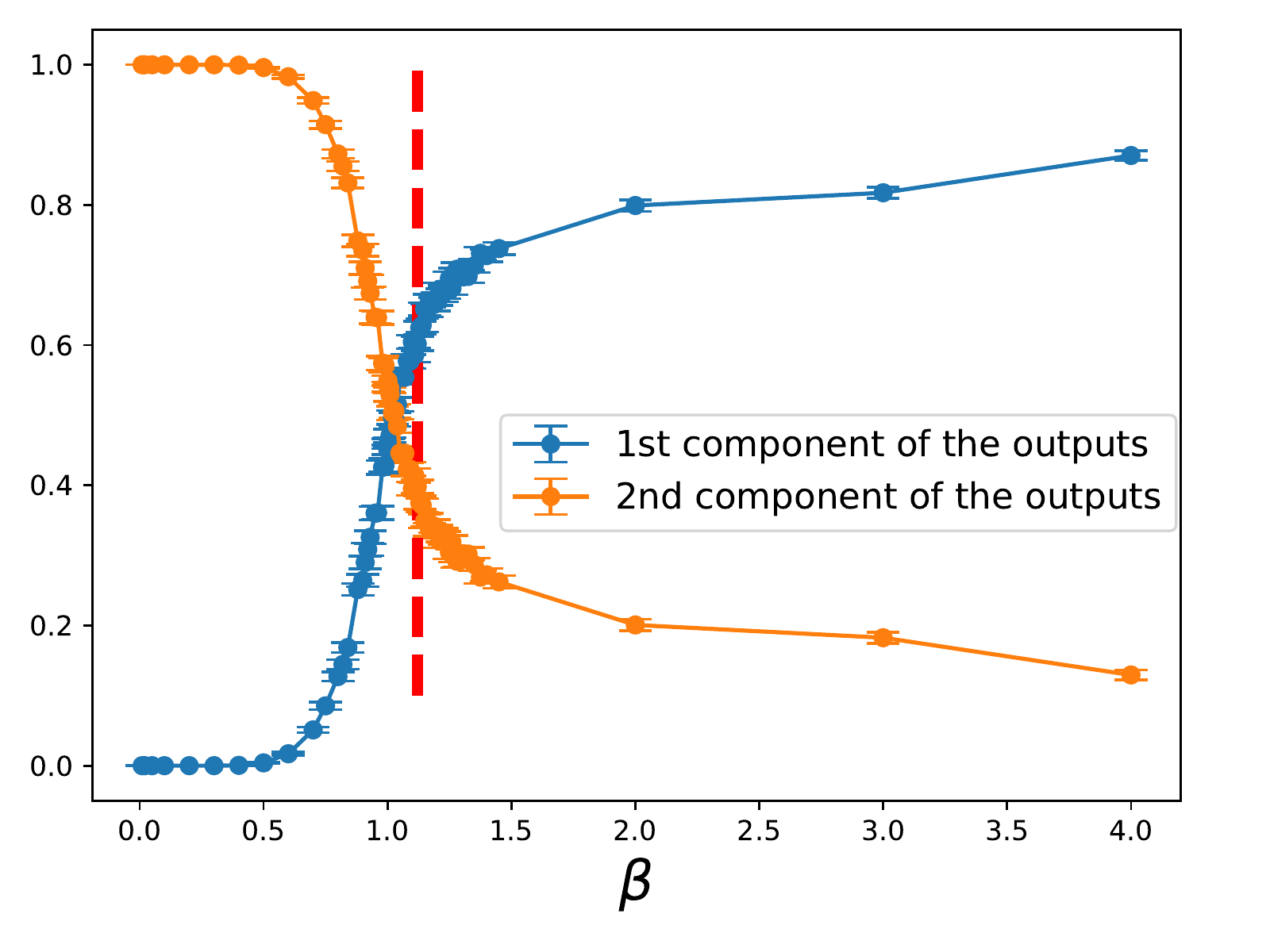}
                  \includegraphics[width=0.315\textwidth]{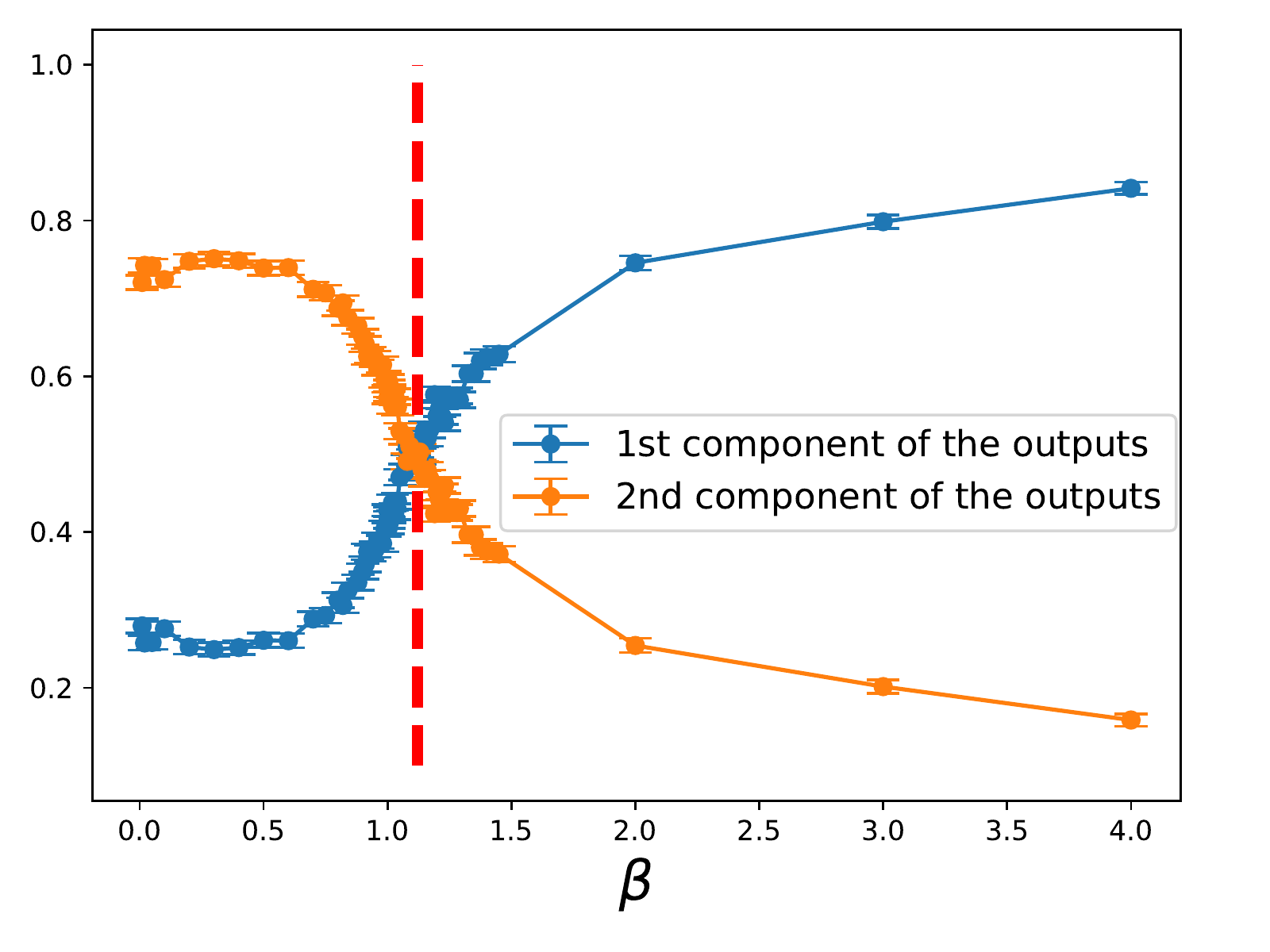}
			\includegraphics[width=0.315\textwidth]{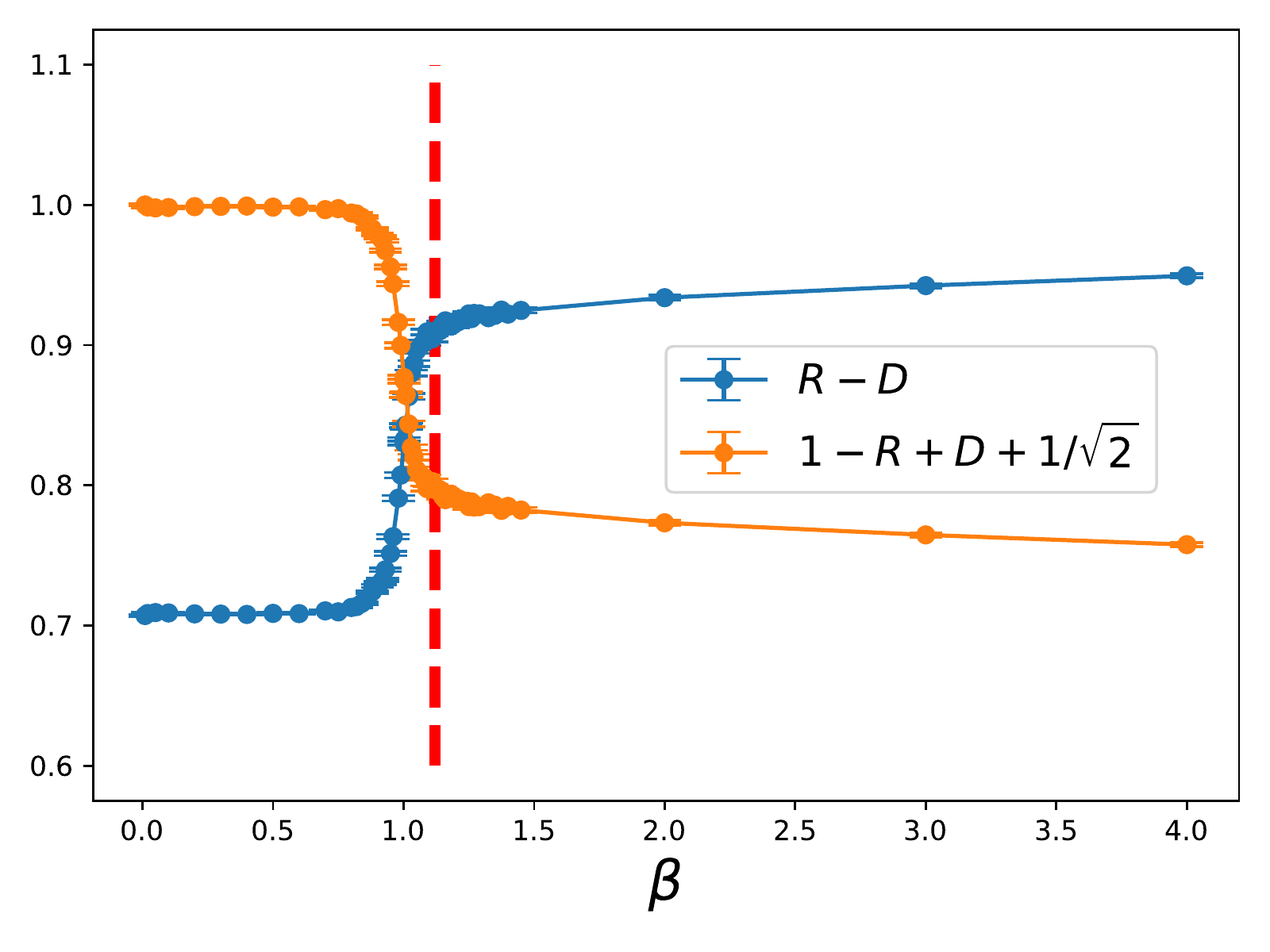}
		}
               \end{center}
	\caption{The CNN determination of the finite size critical inverse temperatures
          for the 2D classical $XY$ model (The system size is $L=128$). The left, the middle, and the
          right panels
          correspond to the first, the second, and the third training methods, respectively, see the main text for the details.
        Here $R$ are the norms of the output vectors and $D$ is the difference between the $R$ of the smallest $\beta$ and $1/\sqrt{2}$.}
	\label{Benchmark_2}
\end{figure}

The results implies that the time required to train a CNN using either the first
or the second
methods is more than 500 times as much as that needed to train a CNN with the third
method.

\begin{figure}
	\begin{center}
		
		\hbox{
		  \includegraphics[width=0.315\textwidth]{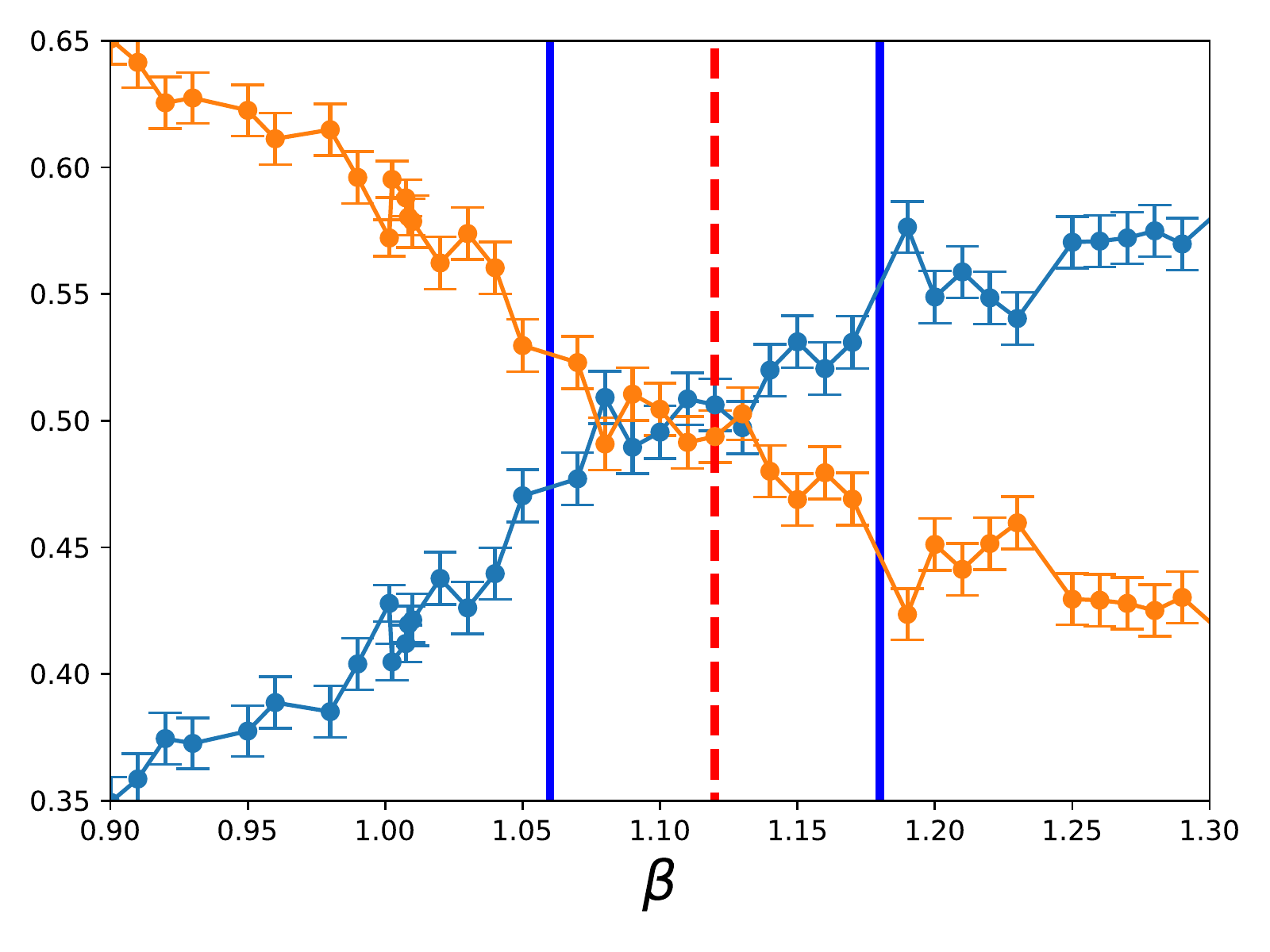}
                  \includegraphics[width=0.315\textwidth]{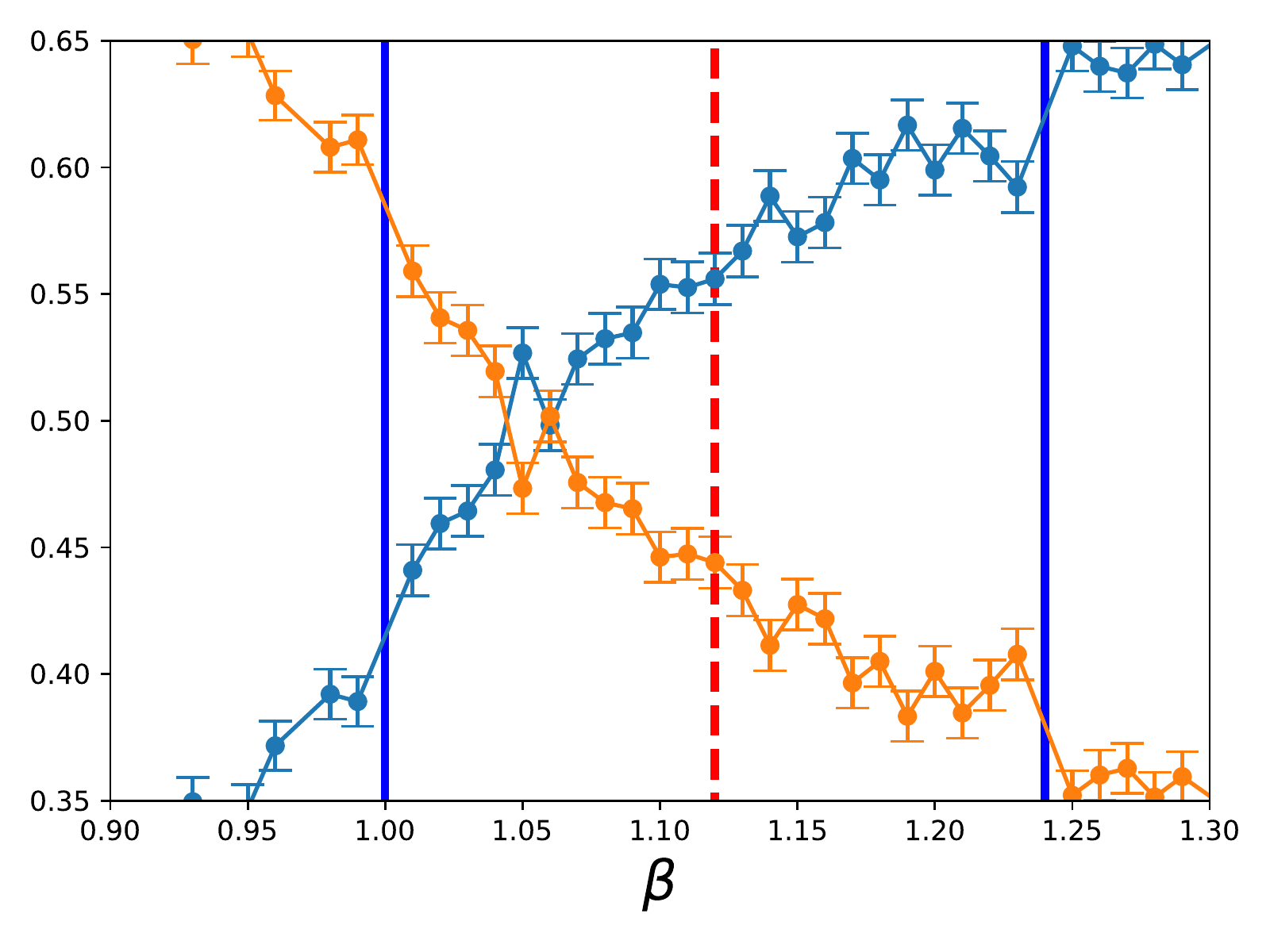}
			\includegraphics[width=0.315\textwidth]{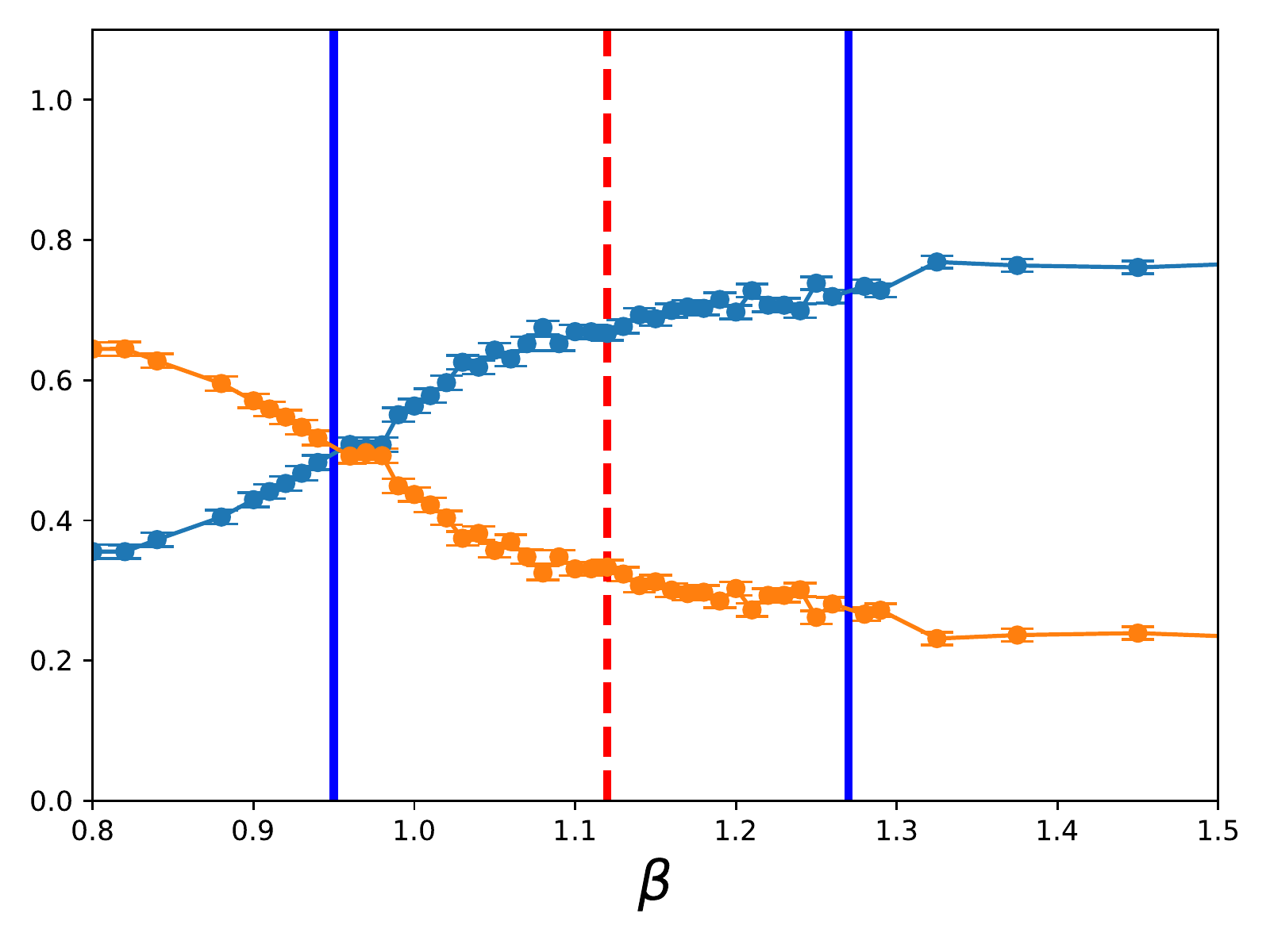}
		}
               \end{center}
	\caption{The CNN determination of the finite size critical inverse temperatures
          for the 2D classical $XY$ model using various training set (The system size is $L=128$). The trainings are conducted with
          the second method.
          In all panels, the vertical dashed lines are the expected $\beta_{\text{KT}}$,
and the intervals bounded by the vertical solid lines are the regions where data associated with them are not used as the training set.
}
	\label{Benchmark_3}
\end{figure}

The CNN predictions are shown in Fig.~\ref{Benchmark_2}.
The left, the middle, and the right panels are the outcomes obtained from the
CNNs trained by the first, the second, and the third method, respectively.
The crossing points and the vertical lines in these panels are the CNN predicted
and the expected critical points, respectively.
Although the second method seems to lead to a result that is closest to $\beta_{\text{KT}}$,
multiple crossing points are found for the left and the middle panels of Fig.~\ref{Benchmark_2}.
This will lead to large uncertainties for the estimated values of $\beta_{\text{KT}}$.  
One can build a more dedicated CNN, consider larger number of epochs, choose smaller learning rate,
or use more data (close to $\beta_{\text{KT}}$) for the training
so that the phenomenon of multiple crossing points does not show up.
However, with these mended strategies the required training time could be (much) longer than that show above.
It should also be pointed out that the crossing point depends on the chosen training set 
and the temperature interval (Which contains the true critical point) where the data associated with it
are not included in the training set, see Fig.~\ref{Benchmark_3}.

\begin{figure}
	\begin{center}
		
		\vbox{
			\includegraphics[width=0.315\textwidth]{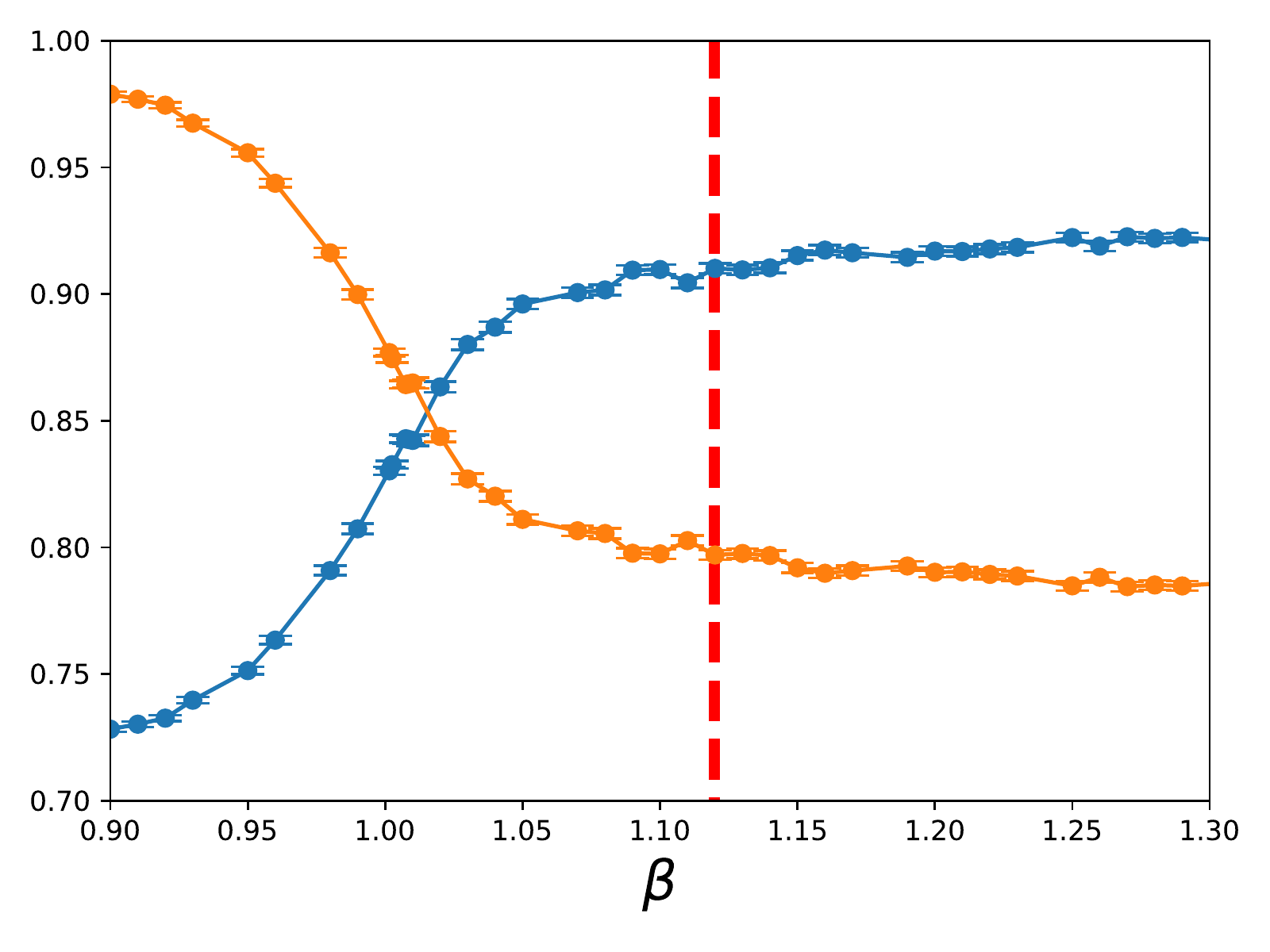}
			\includegraphics[width=0.315\textwidth]{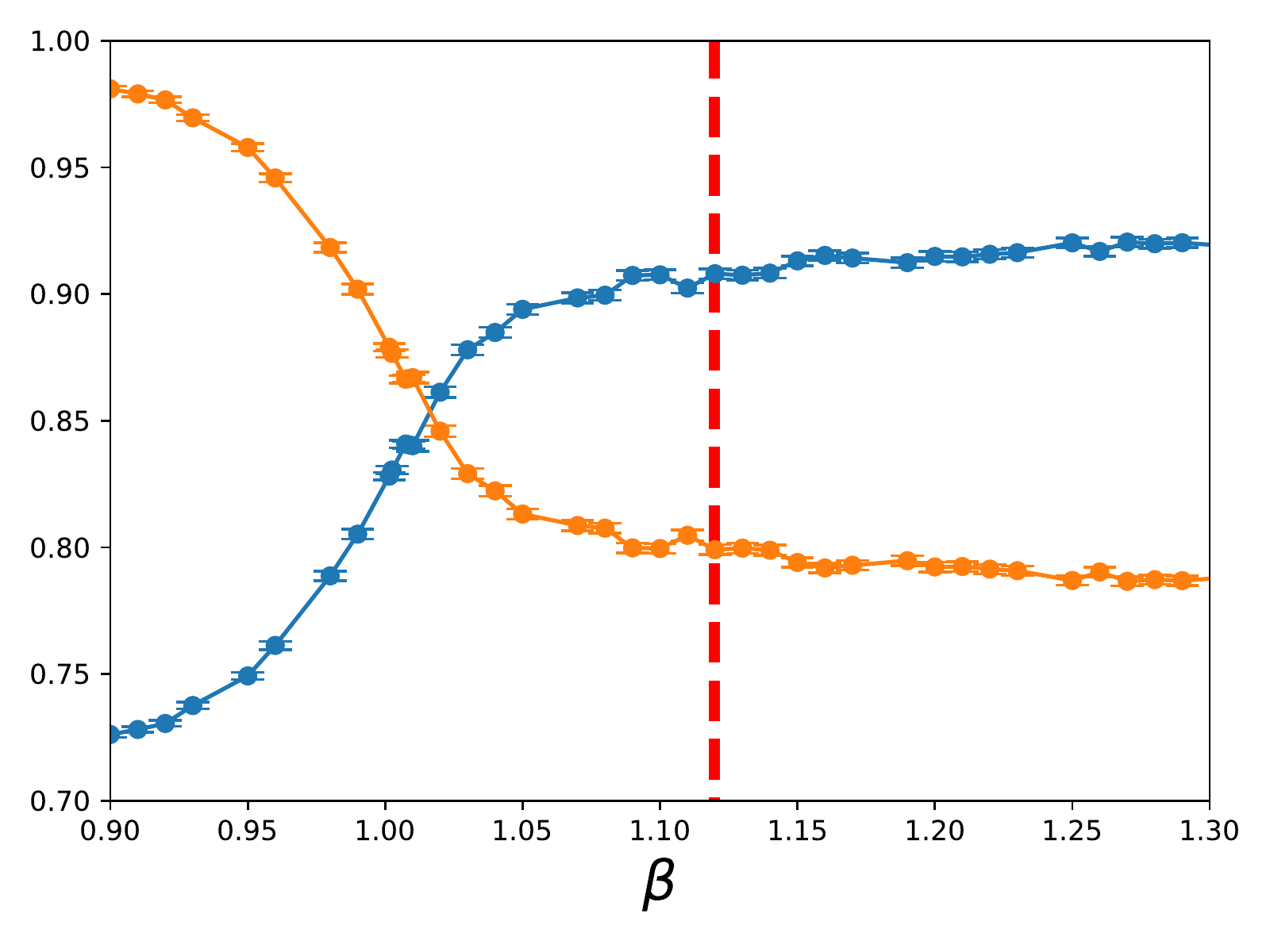}
                        \includegraphics[width=0.315\textwidth]{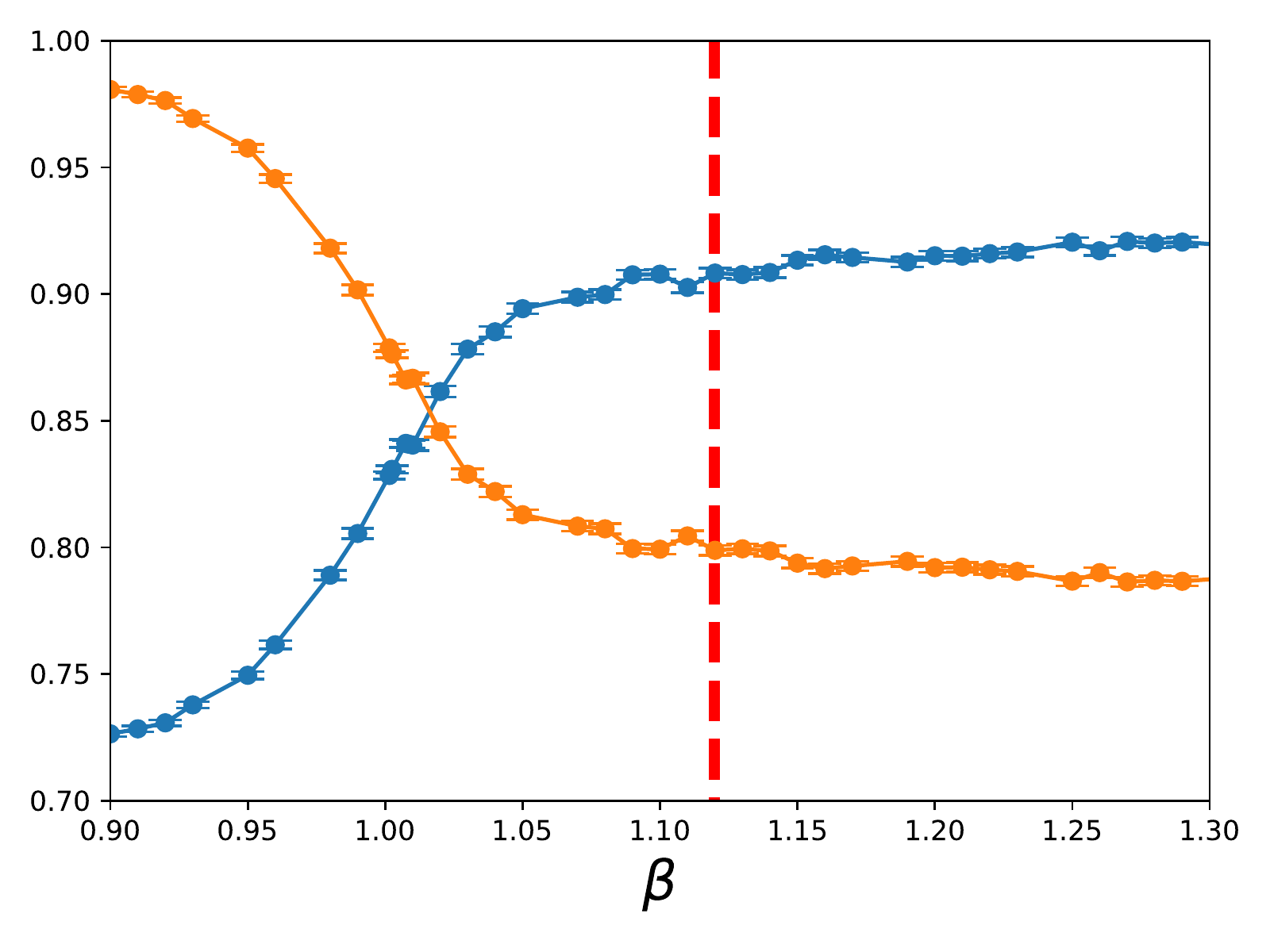}
		}
               \end{center}
	\caption{The CNN determination of the finite size critical inverse temperatures
          for the 2D classical $XY$ model (The system size is $L=128$). The results are associated with
          the third training method and are obtained
          using various $R$ corresponding to different highest temperatures, see the main text for the details.}
	\label{Benchmark_4}
\end{figure}

The determination of the crossing points associated with the third training method involves a factor $D$ which is the difference
between the $R$ of the highest temperature and $1/\sqrt{2}$. As long as the $R$ of low temperatures saturate to a constant,
the resulting crossing point should not depend on which highest temperature data is available (and used), see 
Fig.~\ref{Benchmark_4} which contains the crossings related to the highest temperature being 0.01 (left panel), 0.05 (middle panel),
and 0.1 (right panel).
Finally, although the crossing points in Fig.~\ref{Benchmark_4} seem to be less accurate than that related to the middle panel of
Fig.~\ref{Benchmark_2}, they are of high precision. This will lead to a better determination of the targeted quantity in finite-size
scaling analysis.

\subsection{Comparison with a autoencoder}

The first and the second training approaches shown above require the knowledge of the critical point in advance and this
condition is less desirable since it is not always the case that the critical point is known for any given system.
Hence the NN studies associated with topological phase transitions are usually conducted with unsupervised NNs. Here
we also estimate the time requires to carry out the training for the autoencoder considered in Ref.~\cite{Ale20}.
Due to the infrastructure of the autoencoder built in Ref.~\cite{Ale20}, the raw spin configurations cannot be
employed directly, and we use the first method introduced above (but without considering the step of one-hot encoding) as the training
strategy. 

The total time needed for training the autoencoder is 1432.3 minutes, which is again more than 1000 times as much as that
needed to train a MLP or a CNN with the third (training) method.
In this study we do not conduct a detailed calculation
of the prediction for the constructed autoencoder. A more systematic investigation will be presented in a future work.

\subsection{Remarks}

One can construct a very dedicated and
complicated CNN that has smooth data behavior with respect to
$T$ ($\beta$) and is able to discover the BKT phase transition with high precision.
This would then involve trial and error, and the obtained (complicated) CNN may
only be valid for that studied system.
Even for such a highly engineering CNN, one can still use the 2 artificial configurations introduced above as the
training set, and it is anticipated that the associated comparisons of the efficiency in computation as well as
the accuracy in prediction would be similar to those shown above.

\section{Numerical Results}

In this section the results of applying the 1D NN to calculate the critical
points of the considered models are presented. The associated Monte Carlo simulations are done using the Wolff
algorithm \cite{Wol89}. We would like to emphasize the fact again that NO NEW NN IS TRAINED for obtaining
the main results shown in this study.
First of all, the outcomes associated with the 2D generalized classical $XY$ model are demonstrated.

\subsection{Results associated with the 2D generalized classical $XY$ model}

It is well known that for the 2D generalized classical $XY$ model studied here, two phase transitions will take place as one moves from the low
temperature to the high temperature regions. Moreover, the first and the second phase transitions belong to the 3-state Potts
(with critical temperature $T_{\text{Potts}}\sim 0.365$) and the BKT (with critical temperature $T_{\text{KT}}\sim 0.671$) universality classes,
respectively \cite{Can14,Can16}. 

\subsubsection{Results using the spin configurations}

\begin{figure}
	\begin{center}
		
			\includegraphics[width=0.45\textwidth]{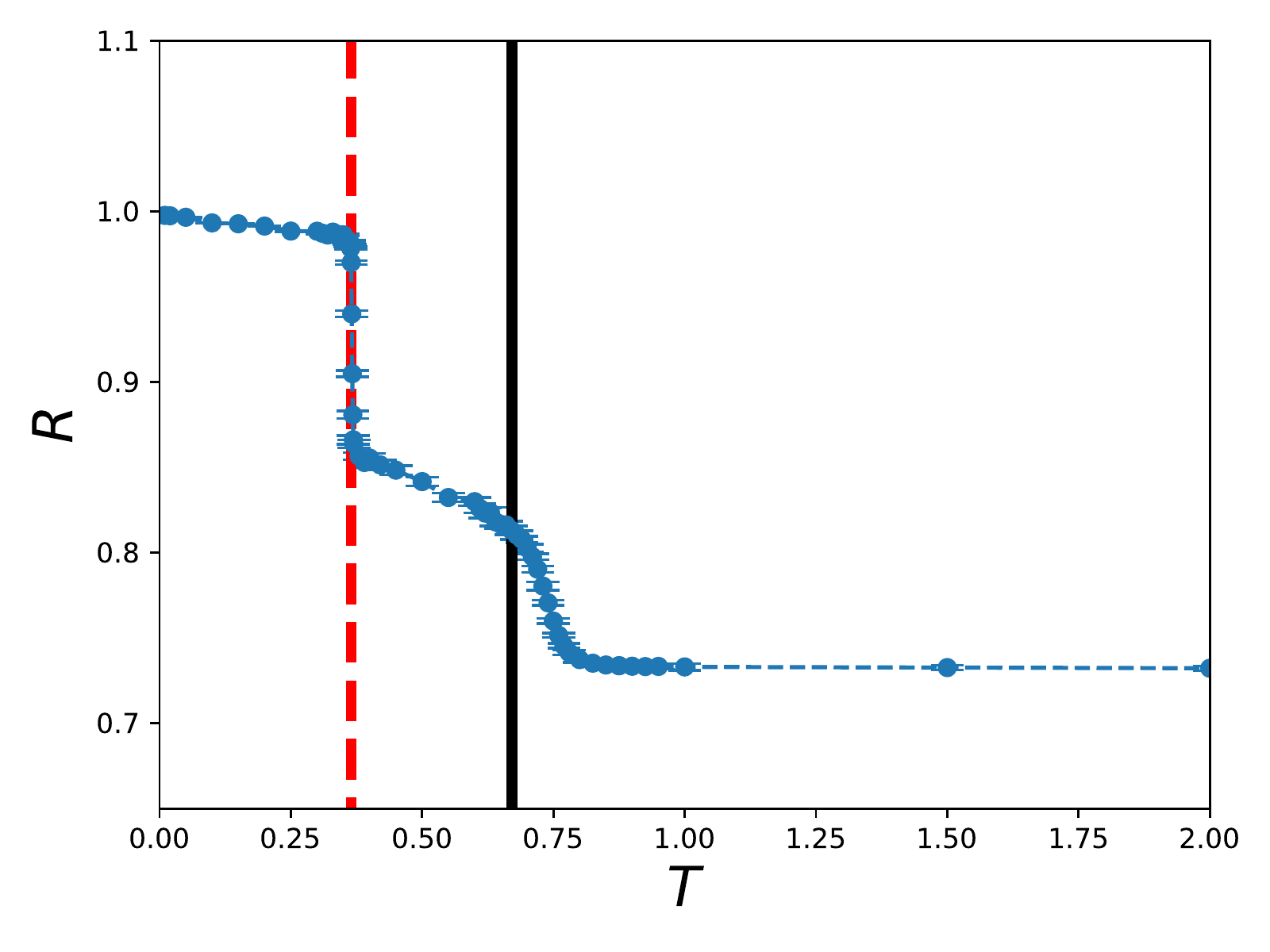}
	\end{center}\vskip-0.7cm
	\caption{The magnitude $R$ of the outputs as a function of
          $T$ for
          the 2D generalized classical $XY$ model (The system size is $L=128$). The outcomes are obtained 
          by using 200 spins ($\theta\,\text{mod}\,\pi$) randomly chosen from each of the real configurations.
        The vertical dashed and solid lines are the expected $T_{\text{Potts}}\sim 0.365$ and $T_{\text{KT}} \sim 0.671$, respectively.}
	\label{R_T}
\end{figure}

To examine whether the employed 1D NN can detect these two phase transitions merely through (very small part of) the associated spin configurations,
for each of the simulated
temperatures and for every stored configuration with $L=128$, we have randomly
chosen 200 sites ${j}$ and have used the $\theta_j\, \text{mod}\, \pi$ of these sites to construct
a 200 sites 1D lattice to feed the NN. The magnitude $R$ of the resulting output vectors, obtained by this mentioned procedure, as a function of $T$
are shown in Fig.~\ref{R_T}.
It should be pointed out that in Fig.~\ref{R_T} the dashed and solid vertical lines represent the expected phase transitions related to the 3-state
Potts and the BKT universality classes, respectively.

Remarkably, as $T$ increases, the temperatures at which $R$ drops significantly are exactly located at those where the phase transitions occur.
This strongly suggests that the used NN, which was trained on 1D lattice of 200 sites, is not only able to find phase transition associated with
spontaneous symmetry breaking, but also can detect BKT type phase transition. It is remarkable as well that the results in Fig.~\ref{R_T} imply
only (very) little information of the system is sufficient to detect change of states of that system.

As we will show later, if the Hamiltonian $H$ and the associated analytic investigation of $H$ are not known in advance, then the
commonly use observables Binder ratios $Q_{i,1}$ can only detect the first phase transition (3-state Potts). Similarly, $Q_{i,3}$ or
the helicity modulus can only find the 2nd transition of BKT type. With our simple NN, even with very LITTLE INFORMATION of the system one 
finds both the transitions. This can be considered as an advantage of NN approach over the traditional methods.

We would like to emphasize the fact that in Fig.~\ref{R_T}, the drop of $R$
near $T_{\text{KT}}$ is slower than that close to $T_{\text{Potts}}$. This is
consistent with the features of symmetry breaking and BKT type transitions since
BKT type transition receives certain logarithmic corrections.
It is remarkable that only few percent of the whole spin configurations
can reveal the characteristics of the associated phase transitions.

\begin{figure}
	\begin{center}
		
		\vbox{
			\includegraphics[width=0.45\textwidth]{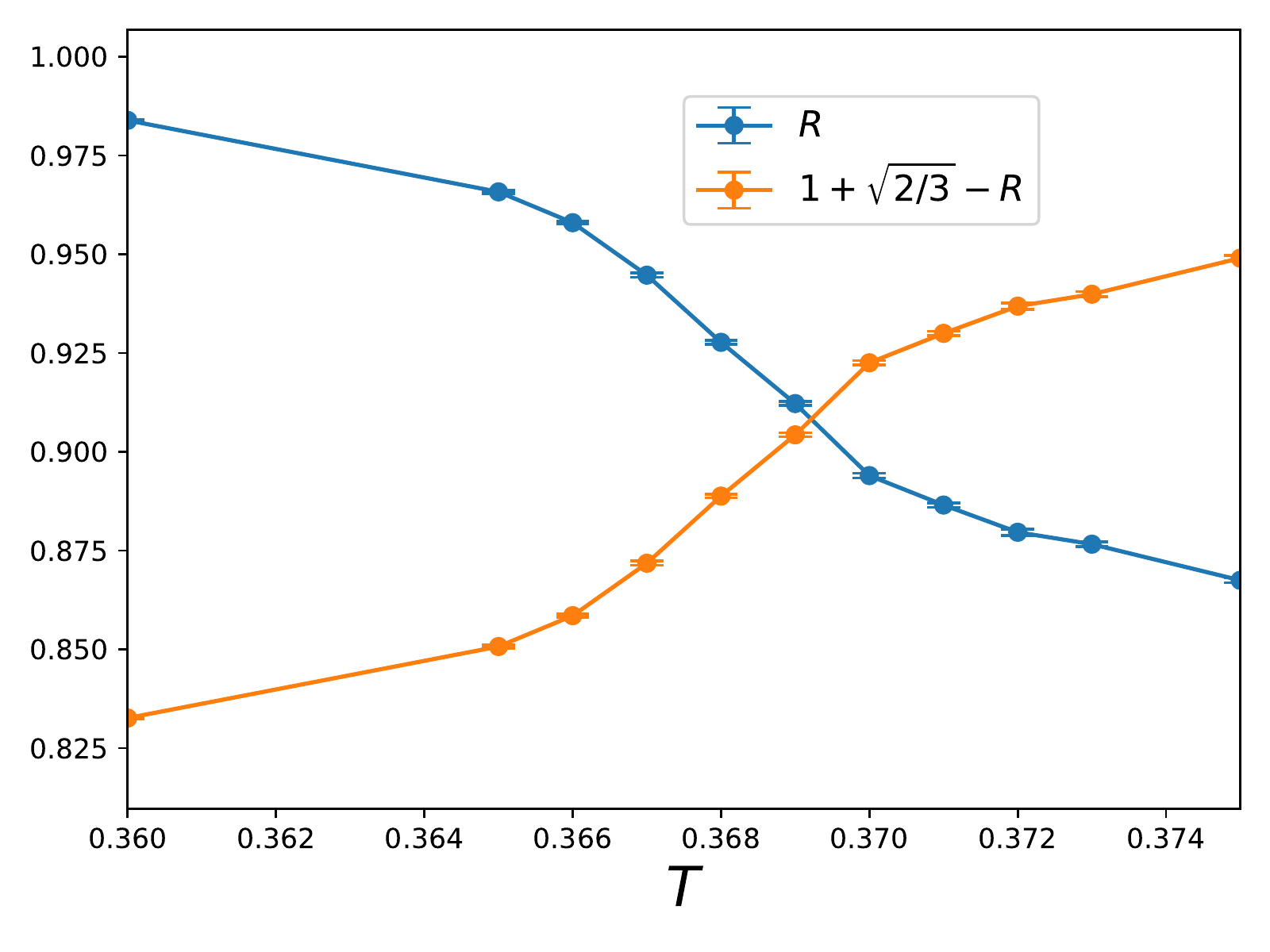}
			\includegraphics[width=0.45\textwidth]{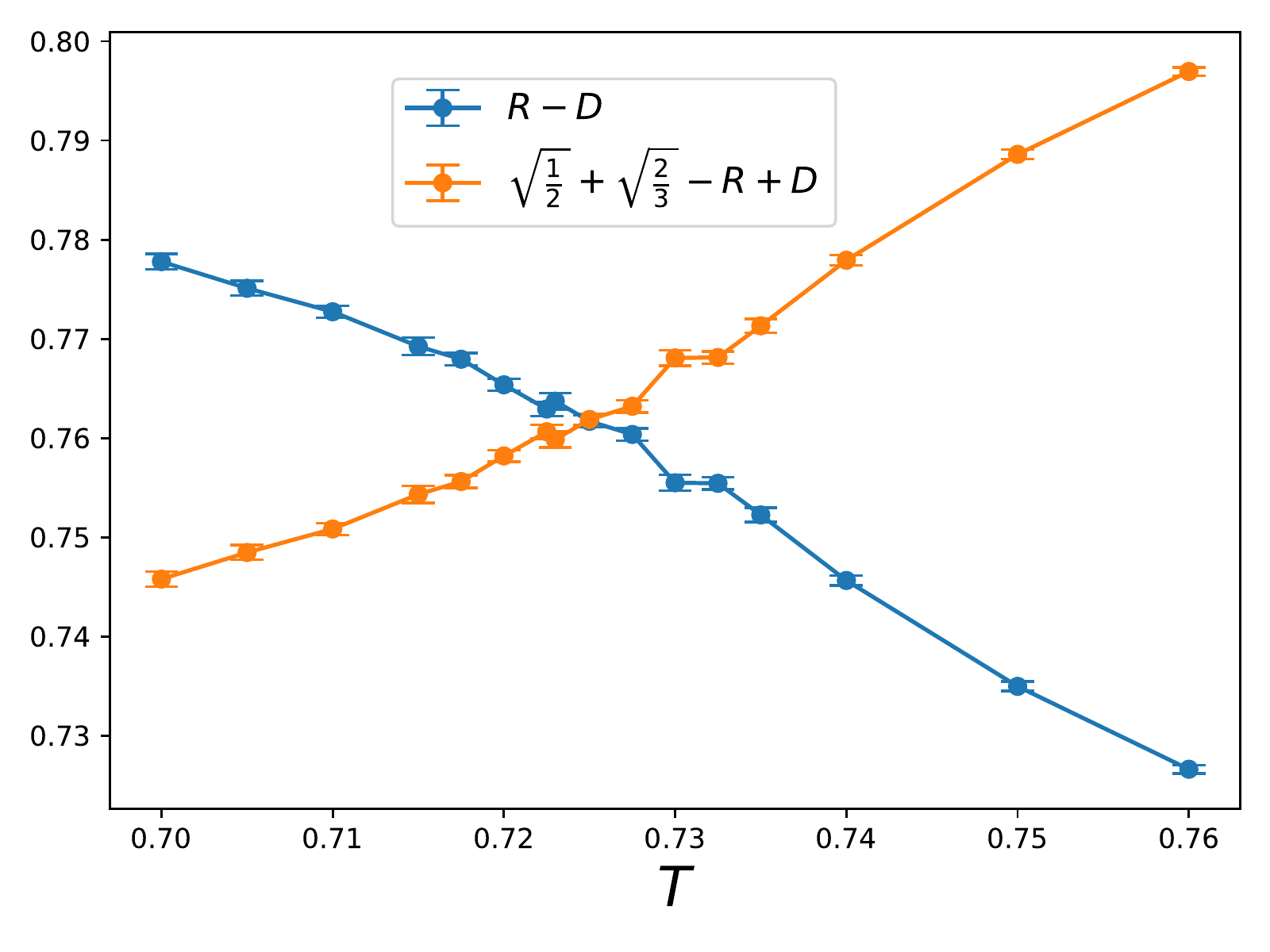}
		}
               \end{center}
	\caption{The determination of the finite size critical temperatures
          for the 2D generalized classical $XY$ model. The left and right panels
          correspond to box sizes $L=64$ (3-state Potts class) and $L=128$ (BKT), respectively.
          The outcomes are obtained
          by using 200 spins ($\theta\,\text{mod}\,\pi$) randomly chosen from each of the real configurations. }
	\label{crossing_Potts_KT}
\end{figure}

To precisely calculate $T_{\text{Potts}}$ and $T_{\text{KT}}$, we have performed
the same procedures for various system sizes. A preliminary investigation
indicates $R \sim 1$ and $R \sim \sqrt{2/3}$ when $T \ll T_{\text{Potts}}$ and
$T_{\text{Potts}} < T \le T_{\text{KT}}$, respectively.
In addition, $R$ $\sim$ $1/\sqrt{2}$ for
$T \gg T_{\text{KT}}$. Using these results and the method introduced in Ref.~\cite{Tan20.2},
the transition temperature $T_{\text{Potts,L}}$ corresponding to a given finite
$L$ can be obtained by the intersection of $R$ and $1+\sqrt{2/3}-R$. Moreover,
For a finite box size $L$ the related $T_{\text{KT,L}}$ is determined by
the crossing point of $R-D$ and $\sqrt{2/3}+1/\sqrt{2}-R+D$, where
$D$ is the difference between the $R$ of the (available) highest temperature
and $1/\sqrt{2}$. The left and right panels of Fig.~\ref{crossing_Potts_KT}
demonstrate the critical temperatures obtained by this procedure for two system
sizes $L=64$ and $L=128$.

\begin{figure}
	\begin{center}
		
		\vbox{
			\includegraphics[width=0.45\textwidth]{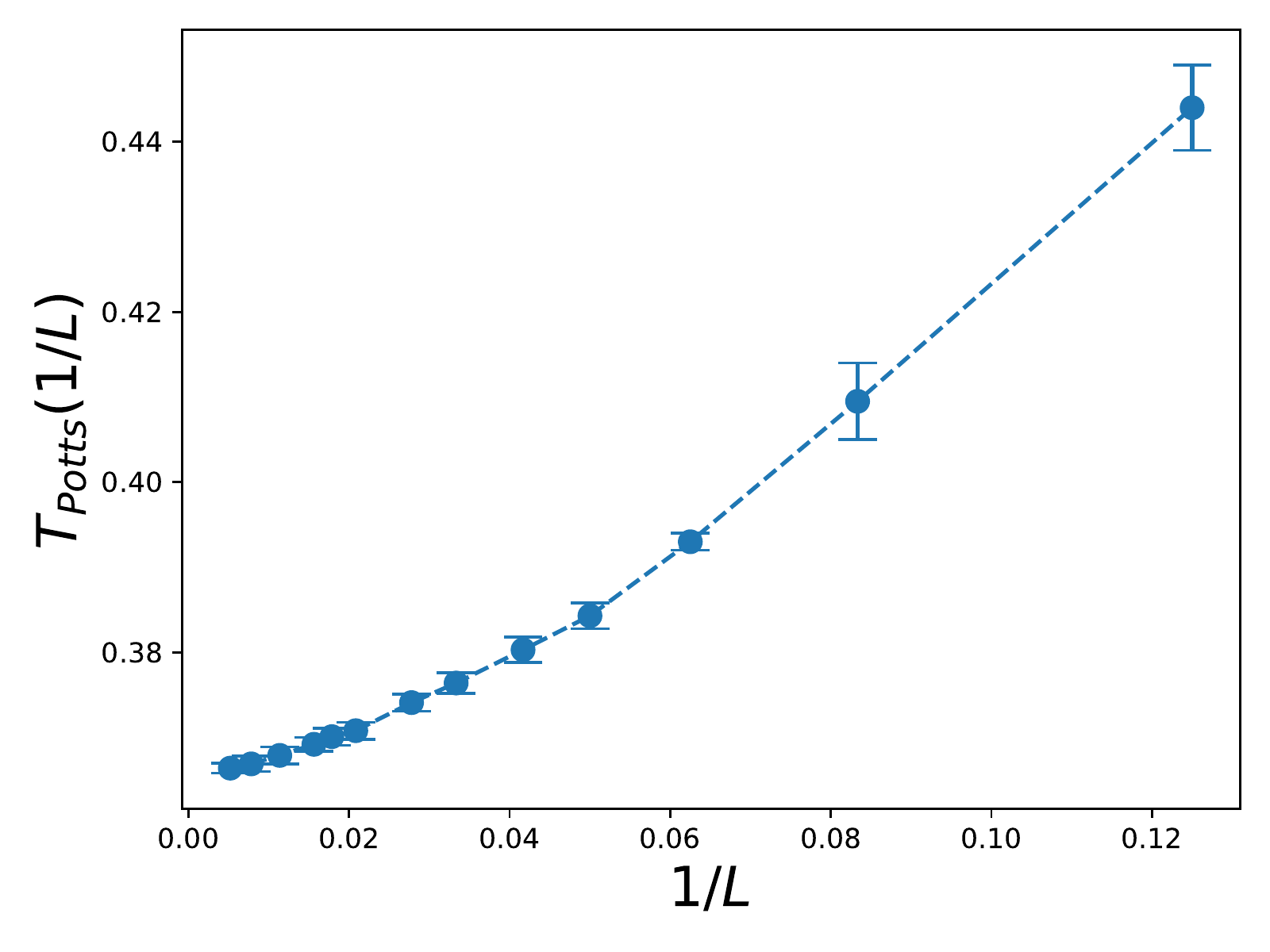}
			\includegraphics[width=0.45\textwidth]{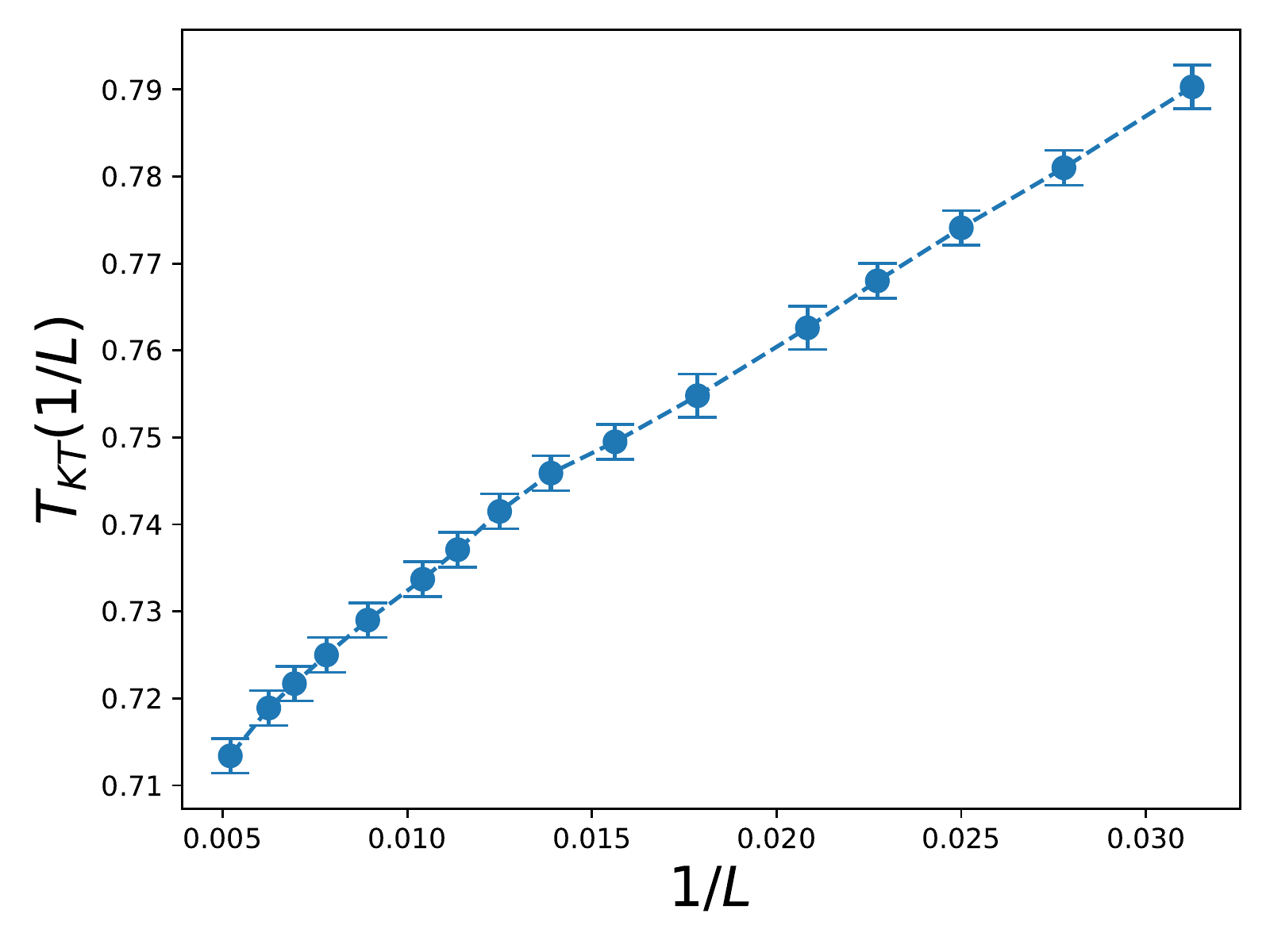}
		}
               \end{center}
	\caption{$T_{\text{Potts,L}}$ (left) and $T_{\text{KT,L}}$ (right) 
          as functions of
          the system size $1/L$ for
          the 2D generalized classical $XY$ model. The outcomes are obtained
          by using 200 spins ($\theta\,\text{mod}\,\pi$) randomly chosen from each of the real configurations. }
	\label{Tc_Potts_KT}
\end{figure}

The intersecting points obtained by the above described strategies associated
with the 3-state Potts and the BKT phase transitions are given as the left and the right
panels of Fig.~\ref{Tc_Potts_KT}.

With a fit of the form $T_{\text{Potts,L}} = T_{\text{Potts}} + 1/L^{\alpha}$ using all the data associated with the 3-state Potts class,
one arrives at $T_{\text{Potts}} = 0.3658(6)$ and $\alpha = 1.52(9)$. The obtained $T_{\text{Potts}} = 0.3658(6)$ agrees well with
$T_{\text{Potts}} \sim 0.365$ determined in Refs.~\cite{Can14,Can16}. Similarly, by fitting all the results related to the BKT transition with the
ansatz $T_{\text{KT,L}} = T_{\text{K}} + b/(\log(L))^2 + c/(\log(L))^4$ ($T_{\text{KT,L}} = T_{\text{K}} + b/(\log(L))^2$), one finds
$T_{\text{KT}} = 0.65(1)$ (0.657(3)) which is slightly below that calculated
in Refs.~\cite{Can14,Can16} (There $T_{\text{KT}}$ is found to be around 0.671). Considering the simplicity of the used NN as well as the
difficulty in the calculations due to the logarithmic corrections associated with the BKT transition, the obtained $T_{\text{KT}} = 0.65(1)$ is
remarkably good.

\subsubsection{Results using the Binder ratios}

In addition to using raw spin configurations for the NN prediction, it is proposed and verified in Ref.~\cite{Tan21} that bulk quantities that saturate to
constants in the low and the high temperature regions can be employed to constructed the needed configurations for the NN prediction. Here we use this
method to reinvestigate the BKT transition of the 2D generalized classical $XY$ model. Notice only several data points are needed to
determine whether the targeted observable is appropriate for performing such calculations.

\begin{figure}
	\begin{center}
			\includegraphics[width=0.45\textwidth]{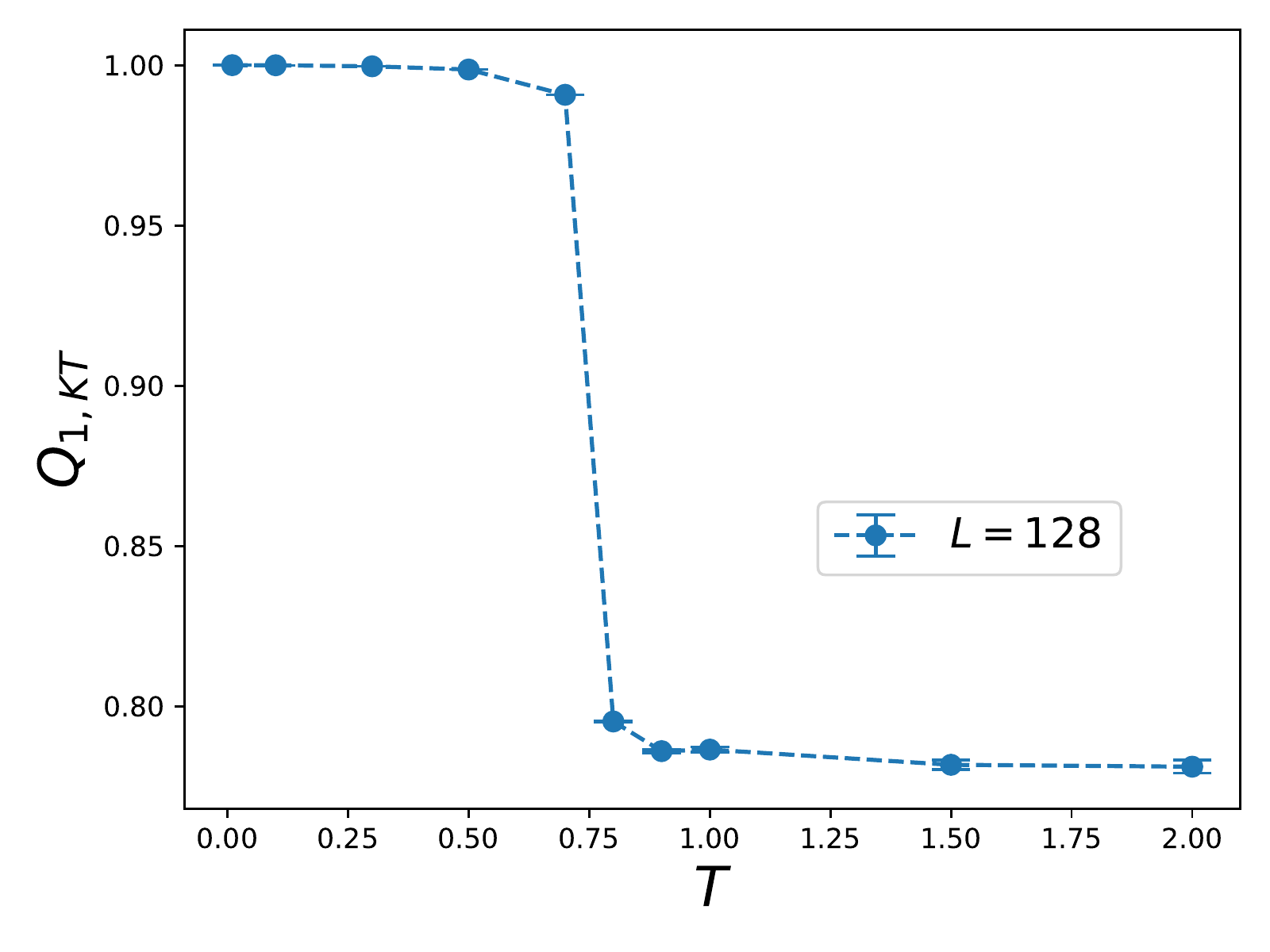}
		
	\end{center}\vskip-0.7cm
	\caption{$Q_{1,3}$ (obtained
		on $128^2$ lattices)
		as a function of $T$ for the 2D generalized classical $XY$ model.}
	\label{GO2_MC}
\end{figure}

As can be seen in
Fig.~\ref{GO2_MC}, the first Binder ratio $Q_{1,3}$ fulfills the condition of saturation very well. Consequently, $Q_{1,3}$ is perfectly suitable to
carry out the associated calculations.
It is interesting to notice that in Fig.~\ref{GO2_MC} one sees that the magnitude of $Q_{1,3}$ remains a constant near $T_{\text{Potts}}$.
In other words, $Q_{1,3}$ cannot be used to detect the phase transition associated with the 3-state Potts universality class.
Similar situation occurs for $Q_{1,1}$. These observations in conjunction with Fig.~\ref{R_T} shown in the previous section
strongly suggest that the NN approach has great potential to perform better than the traditional methods when phase transition is concerned.

In our study, the Binder ratio $Q_{1,3} $ is used to
construct the required configurations of 1D lattice with 200 sites
for the prediction. For completeness, the detailed procedures are summarized as follows \cite{Tan21}. 
First, one performs the simulations with different $L$ and $T$. Second, for each simulated $L$, let $Q_{1,3}$
  at the highest and the lowest temperatures be denoted by $Q_{\text{H}}$ and $Q_{\text{L}}$, respectively. In addition, $D_L$ is defined as $D_L = Q_{\text{L}}-Q_{\text{H}}$. With these notations, at every other temperature
a configuration on a 1D lattice of 200 sites which will be used for the prediction
is constructed through the following steps.
\begin{enumerate}
	\item{For a given temperature $T$, let the difference between the $Q_{1,3}$ at $T$ and $Q_{\text{H}}$
		be $D_{T}$.}
	\item{For each site $i$ of the 200 sites 1D lattice, choose a number $n$ in [0,1) randomly and uniformly.}
	\item{If $n \ge |D_{T}/D_{L}|$, then the site $i$ is assigned the integer $1$. Otherwise the integer $0$ is given to $i$. }
	\item{Repeat steps 1, 2 and 3 for all the simulated temperatures (or the inverse temperatures $\beta$).}
\end{enumerate}

  The configurations constructed from steps 1 to 4 introduced above are fed to the trained NN. 
  Then the associated critical point can be calculated by studying the
  temperature (or $\beta$) dependence of the output vectors $\vec{V}$.
  More specifically, if one treats
$\vec{V}$ as functions of $T$ (or $\beta$), the critical point
can be estimated to be at the temperature (or $\beta$) corresponding to the crossing point of the two curves built up from the
components of $\vec{V}$. Apart from this, the critical point can be determined to be the temperature (or $\beta$) at which the output
vector has the smallest value of magnitude as well.

We would like to point out that since with this method only bulk quantities satisfying certain conditions are needed for the NN prediction,
the necessary storage space is much less than that for the standard approaches which require the whole spin states.

\begin{figure}
	\begin{center}
		\vbox{
			\includegraphics[width=0.45\textwidth]{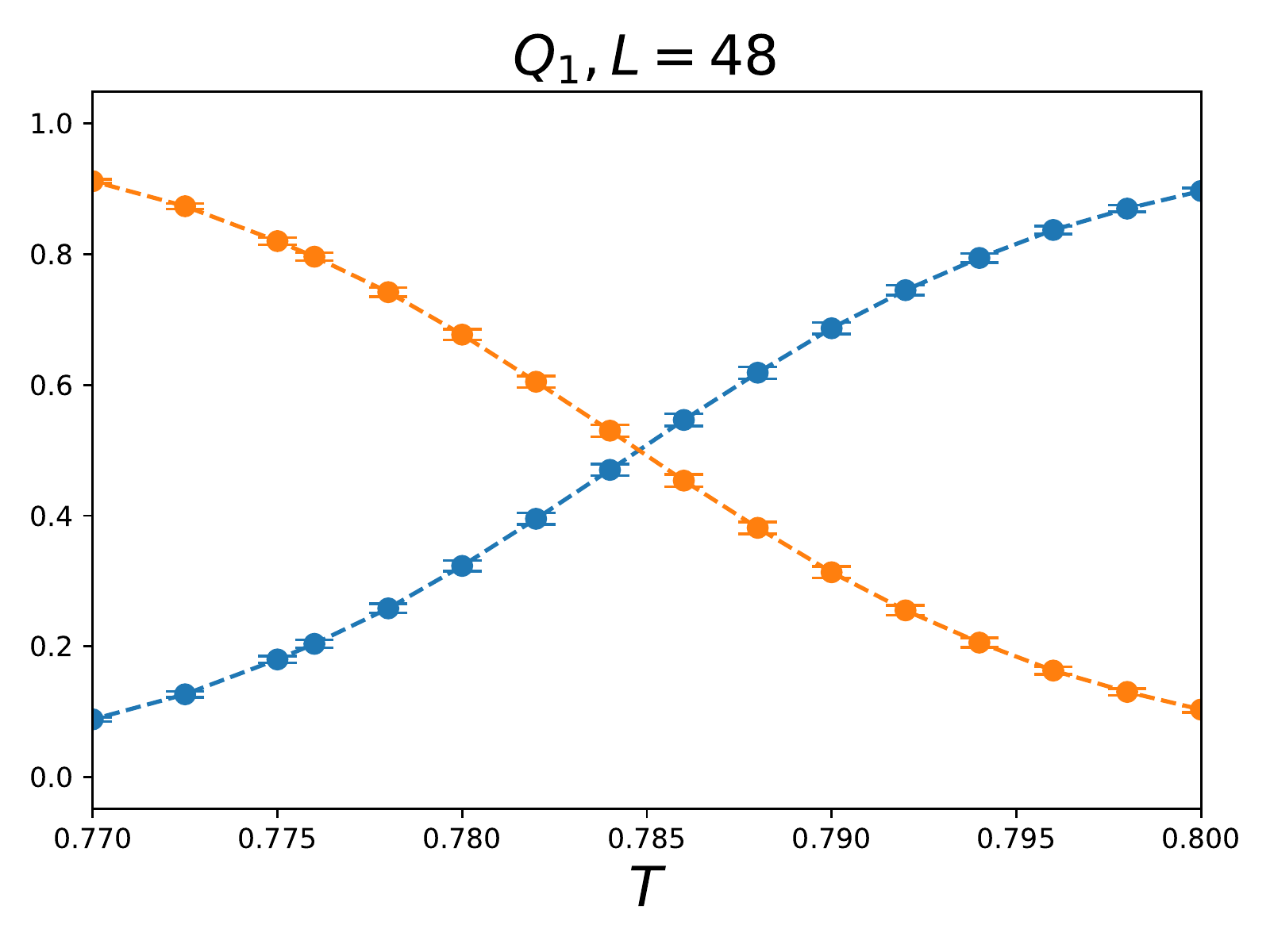}
			\includegraphics[width=0.45\textwidth]{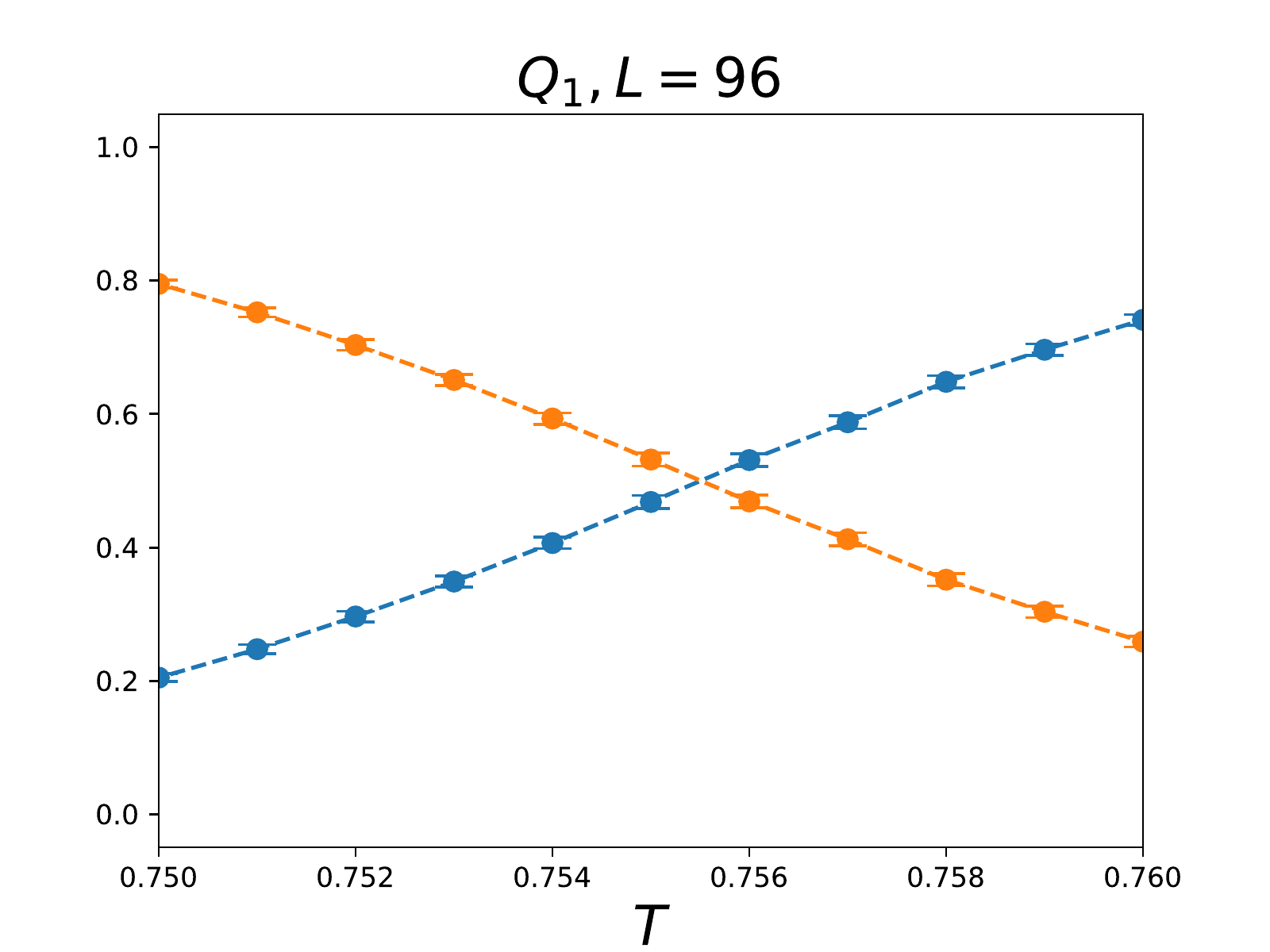}
		}
	\end{center}\vskip-0.7cm
	\caption{The intersecting points of the two components of the output vectors. The associated bulk quantity is $Q_1$. The left and the right
          panels are for the 2D generalized classical
          $XY$ model of system sizes $L=48$ and 96, respectively. }
	\label{crossing_GXY}
\end{figure}

\begin{figure}
	\begin{center}
		
			\includegraphics[width=0.45\textwidth]{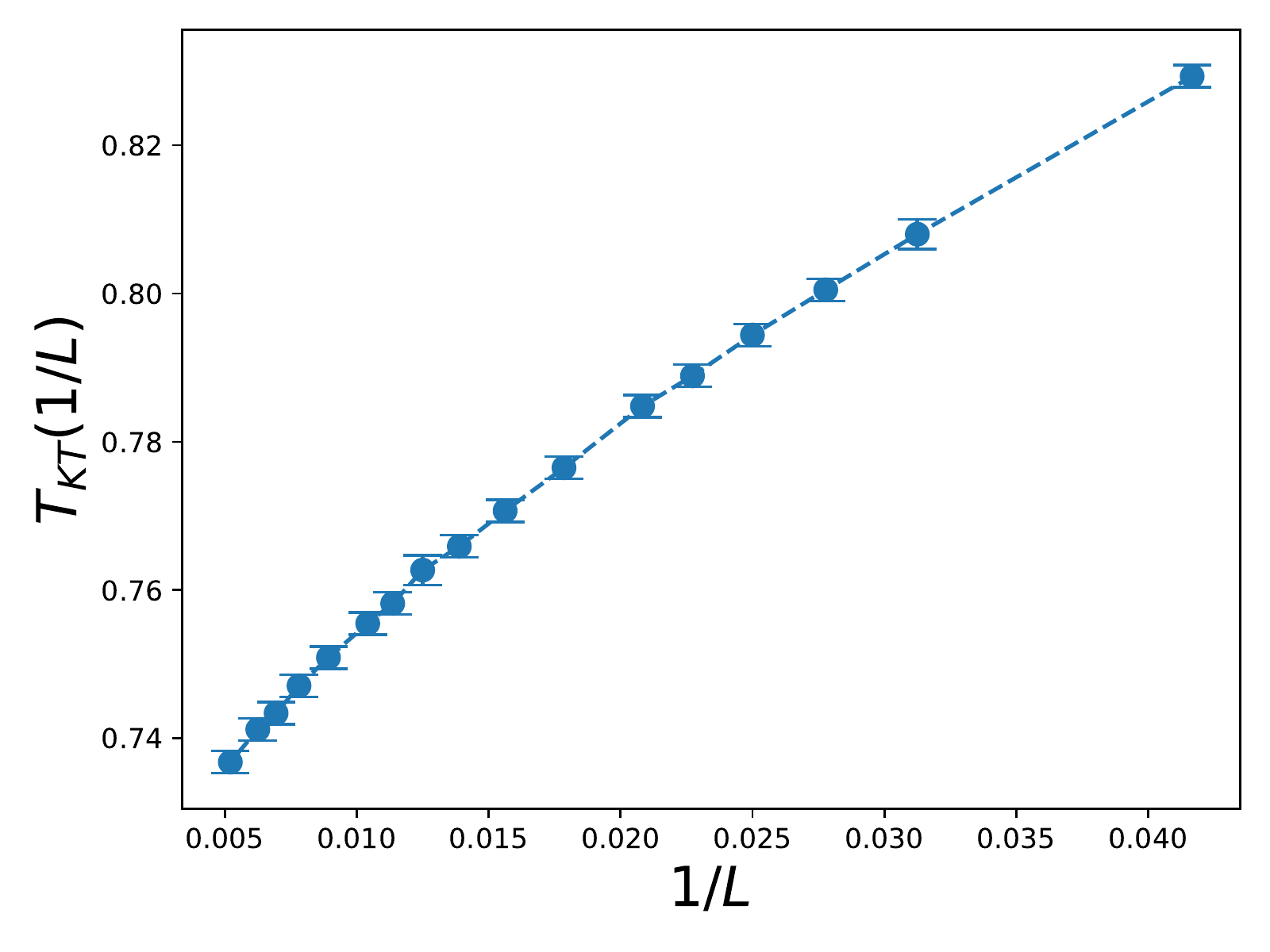}
	\end{center}\vskip-0.7cm
	\caption{$T_{\text{KT}}$ (Which are obtained by the intersection method)
          as a function of the system size $1/L$ for
          the 2D generalized classical $XY$ model. The outcomes are obtained by the intersection method and are based on $Q_1$. }
	\label{Tc_GXY2}
\end{figure}

The intersections obtained through the steps introduced above are shown in
Fig.~\ref{crossing_GXY}. The employed bulk observable for reaching that
figure is $Q_1$ and the system sizes $L$ for the left and the right panels 
of Fig.~\ref{crossing_GXY} are 48 and 96, respectively. In addition, the
crossing points as a function of $1/L$ is demonstrated in Fig.~\ref{Tc_GXY2}.

By applying the finite-size scaling of the expression
$T_{\text{KT,L}} = T_{\text{KT}} + b/(\log(L))^2 + c/(\log(L))^4$ ($T_{\text{KT,L}} = T_{\text{KT}} + b/(\log(L))^2$) to all the data ($L \ge 96$ data) in
Fig.~\ref{Tc_GXY2}, we arrive at $T_{\text{KT}} = 0.674(5)$ ($T_{\text{KT}} = 0.679(7)$) which is in
excellent agreement with $T_{\text{KT}} \sim 0.671$ estimated in Refs.~\cite{Can14,Can16}.

It is interesting to notice that the result calculated by using $Q_1$ seems
more accurate than that obtained with part of the spin configurations.
This may be due to the fact that $Q_1$ is a global
quantity hence reveals more information of the system than that from
part of the raw spin configurations.

\subsection{Results associated with the 2D classical $XY$ model}

\begin{figure}
	\begin{center}
		\vbox{
			\includegraphics[width=0.45\textwidth]{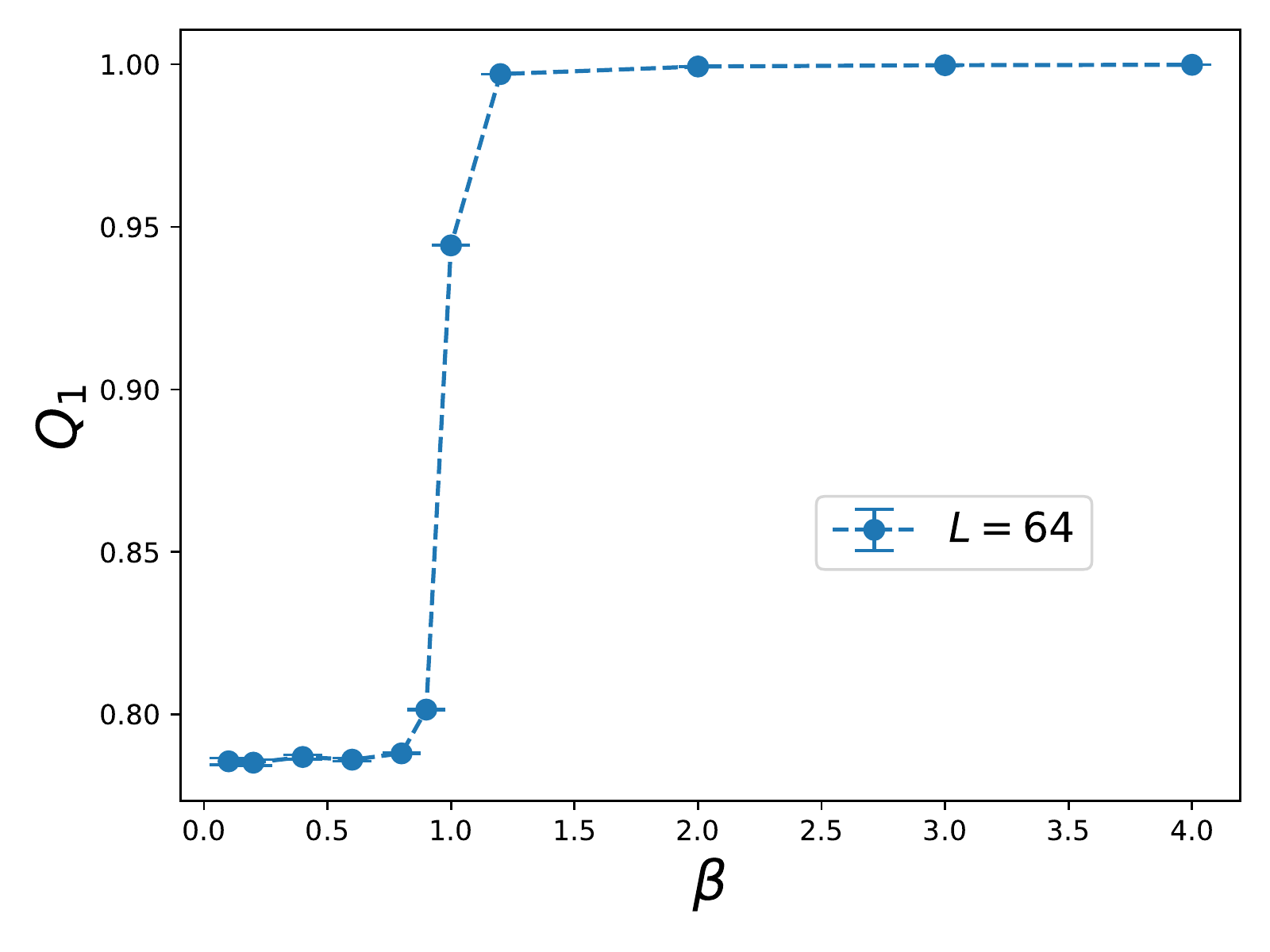}
			\includegraphics[width=0.45\textwidth]{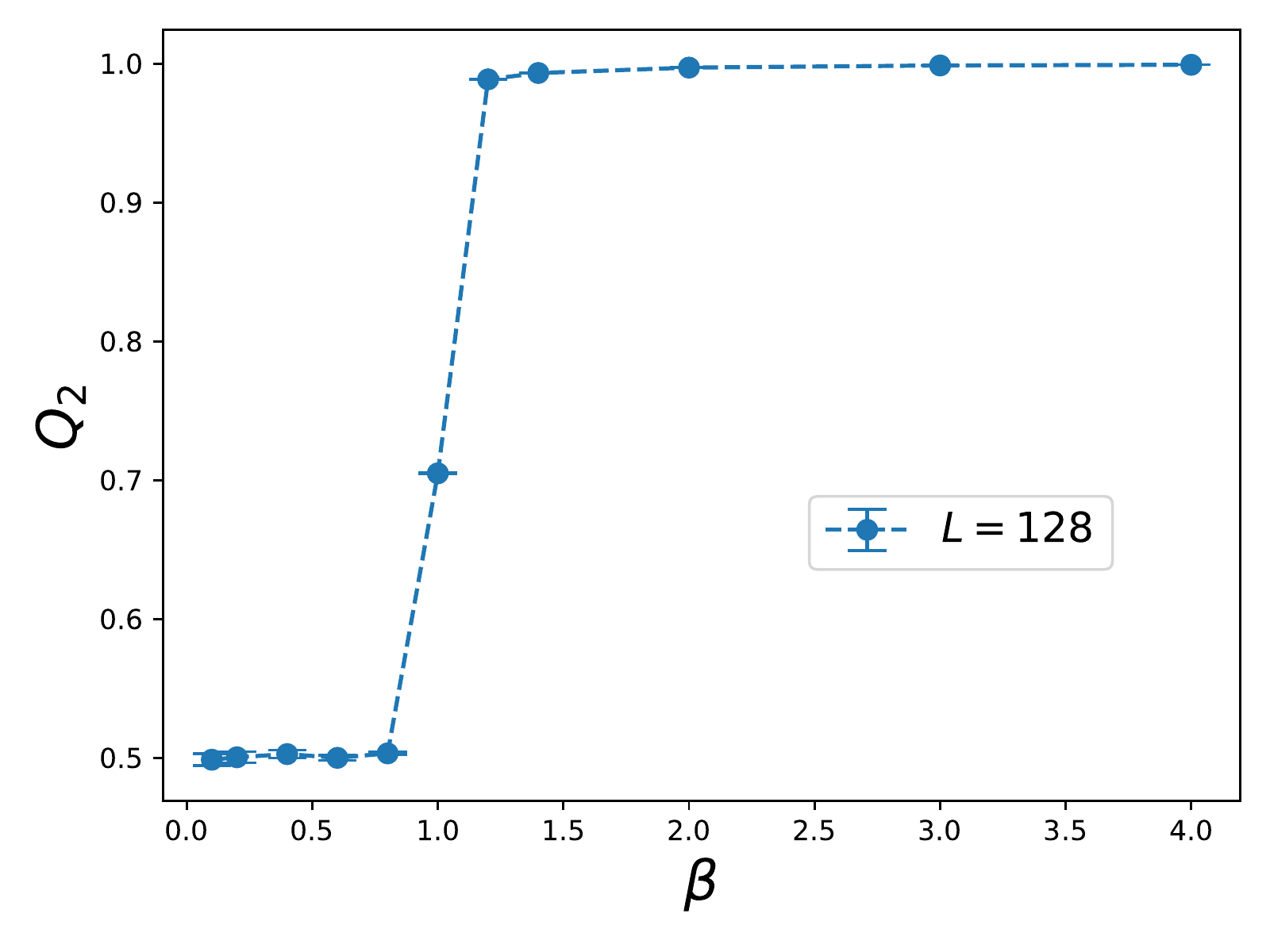}
		}
	\end{center}\vskip-0.7cm
	\caption{$Q_1$ (left panel, obtained on $64^2$ lattices) and $Q_2$ (right panel, obtained
		on $128^2$ lattices)
		as functions of $\beta$ for the 2D classical $XY$ model.}
	\label{O2_MC}
\end{figure}

For the 2D classical $XY$ model, we will consider the approach of using the
bulk observable(s) to determine the associated $T_{\text{KT}}$. Similar to the case of 2D generalized classical $XY$ model,
the first and 2nd Binder ratios fulfill the saturation criterion, see
Fig.~\ref{O2_MC}. Therefore $Q_1$ will be employed for the calculations.

\begin{figure}
	\begin{center}
		\vbox{
			\includegraphics[width=0.45\textwidth]{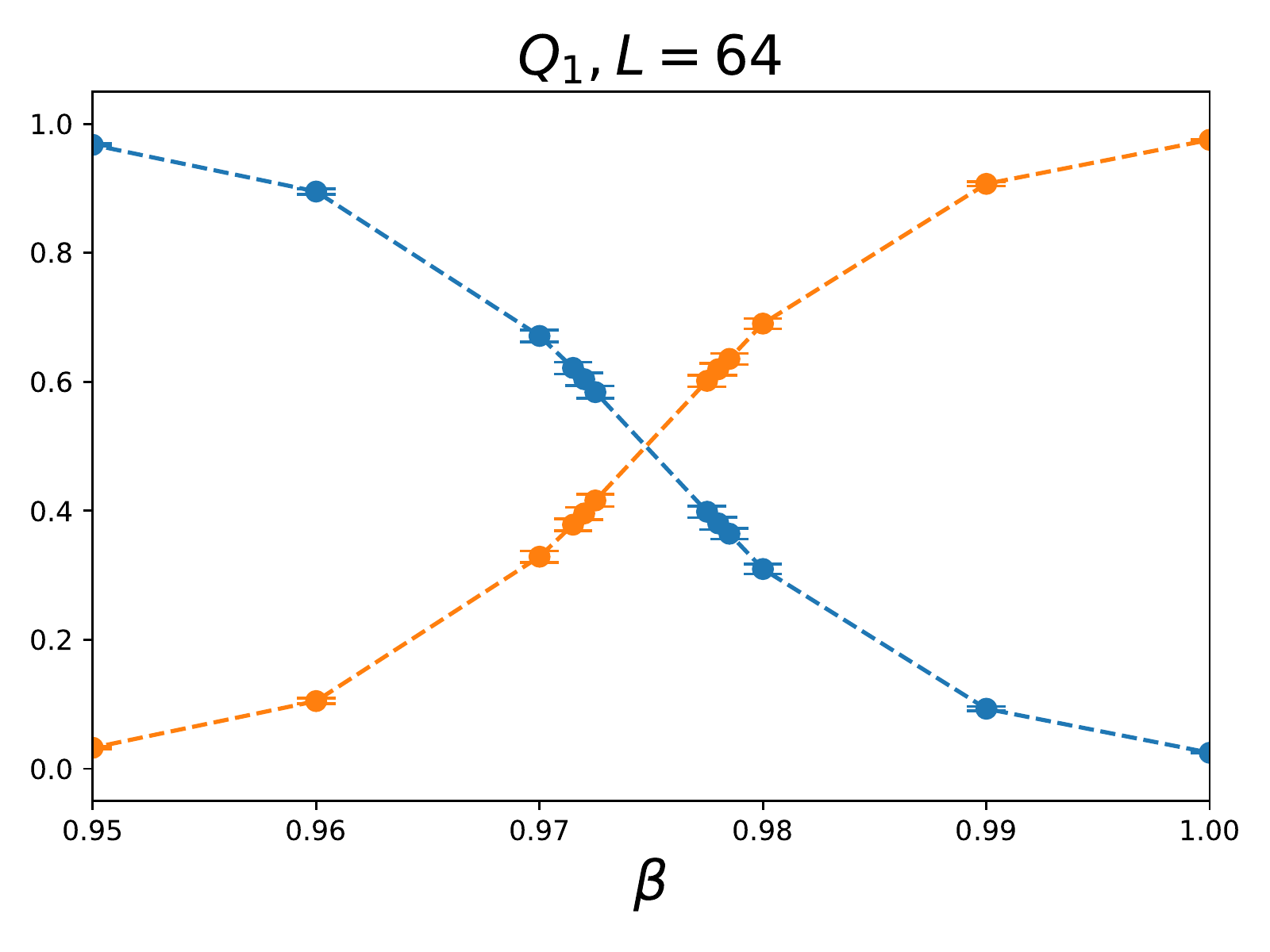}
			\includegraphics[width=0.45\textwidth]{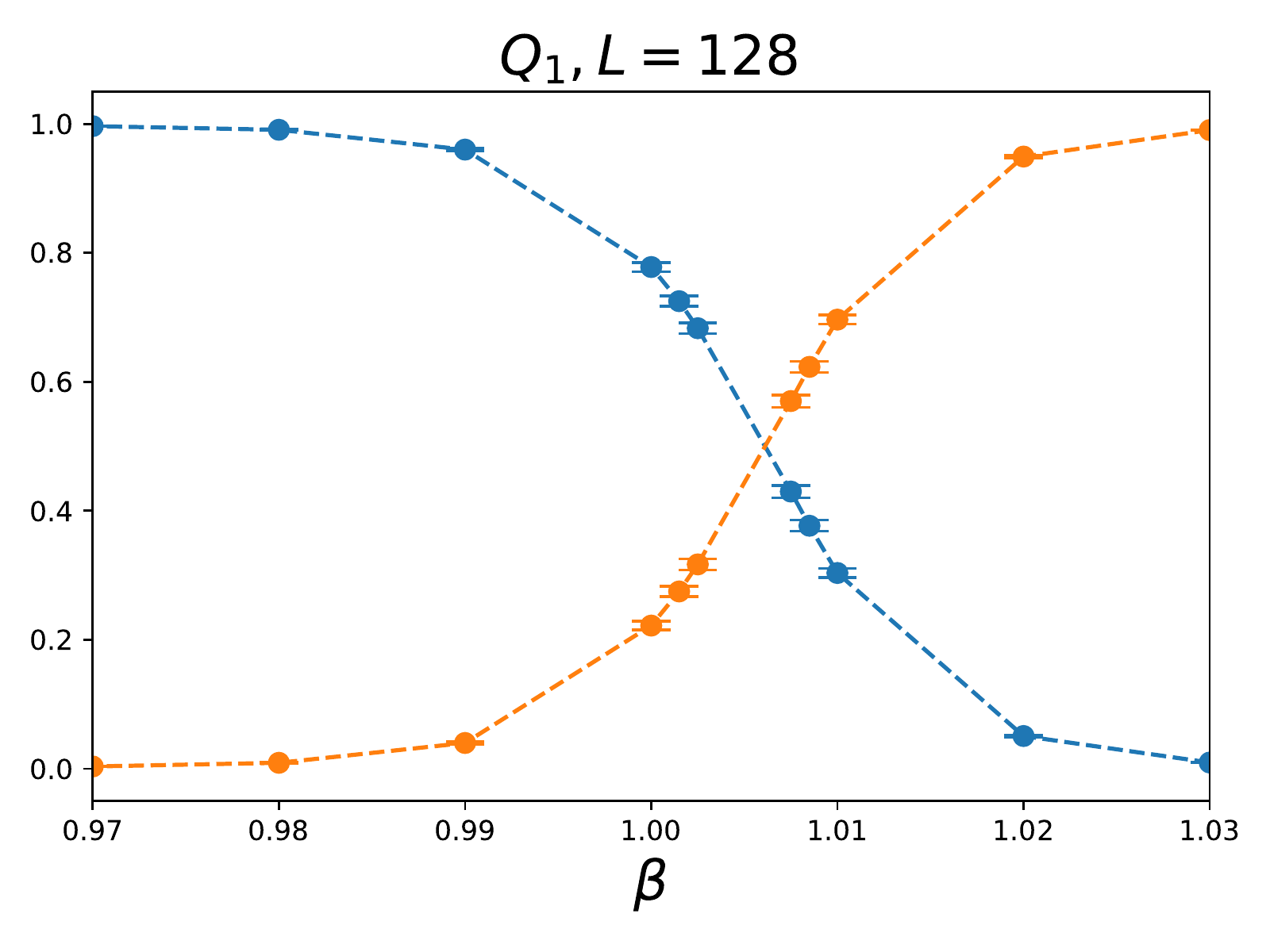}
		}
	\end{center}\vskip-0.7cm
	\caption{The intersecting points of the two components of the output vectors. The associated bulk quantity is $Q_1$. The left and the right panels are for the 2D classical $XY$ model of system sizes $L=64$ and $L = 128$, respectively. }
	\label{crossing_XY}
\end{figure}

Fig.~\ref{crossing_XY} shows the intersections of the two components of
the output vectors. The left and the right panels are for $L=64$ and $L=128$,
respectively.

\begin{figure}
	\begin{center}
		
			\includegraphics[width=0.45\textwidth]{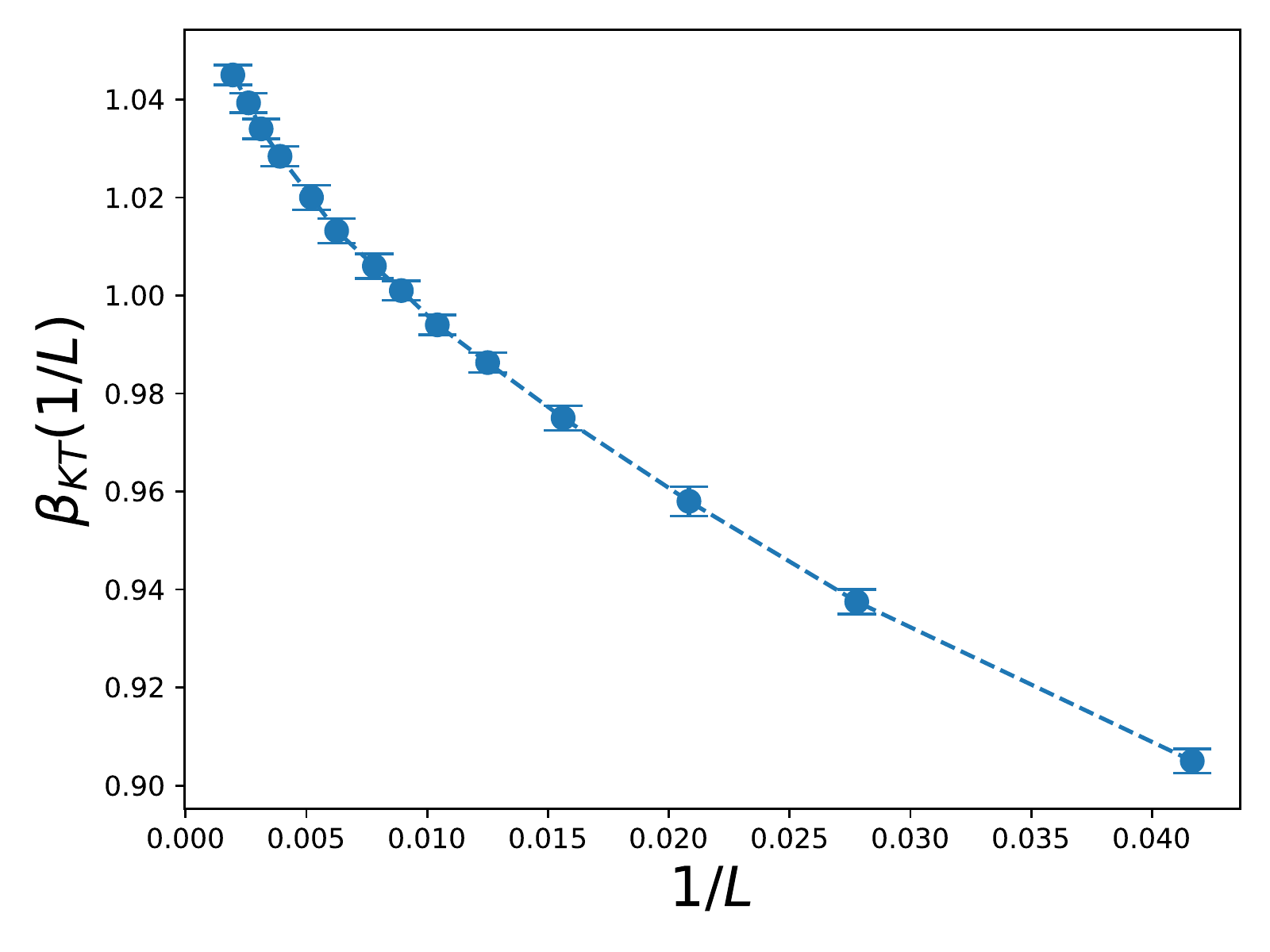}
	\end{center}\vskip-0.7cm
	\caption{$\beta_{\text{KT}}$ as a function of the system size $1/L$ for
          the 2D classical $XY$ model. The outcomes are obtained by the intersection method and are based on $Q_1$. }
	\label{Tc_XY}
\end{figure}

The finite system inverse critical temperatures $\beta_{\text{KT,L}}$ of the
2D classical $XY$ model determined by the intersection idea is
demonstrated in Fig.~\ref{Tc_XY}. Moreover, fits with the ansatz
$\beta_{\text{KT,L}} = \beta_{\text{KT}} + b/(\log(L))^2 + c/(\log(L))^4$ ($\beta_{\text{KT,L}} = \beta_{\text{KT}} + b/(\log(L))^2$) can lead
to $\beta_{\text{KT}} = 1.112(6)$ ($\beta_{\text{KT}} = 1.107(7)$). The obtained values of $\beta_{\text{KT}}$ are in good
agreement with $\beta_{\text{KT}} \sim 1.1199$ claimed in Ref.~\cite{Has05}.
During the analysis, we find that the values of $\beta_{\text{KT}}$ increase
as fewer and fewer data of small $L$ are included in the fits.
This indicates larger $L$ data may be needed in order to reach a high precision agreement.

\section{Conclusions and Discussions}

Using the NN constructed in Ref.~\cite{Tan21}, which was trained on a 1D lattice of 200 sites, we calculate the critical points
of the 2D classical $XY$ and the 2D generalized classical $XY$ models. We would like to stress again the following points.

\begin{enumerate}
\item{In this study, no NN is trained. The employed NN is adopted from Ref.~\cite{Tan21}, which was trained on a 1D lattice of
  200 sites using two artificial configurations as the training set. In other words, ONE SIMPLE
  NN can be considered for the investigations of phase transitions of many
  models, and it is not required to train a new NN whenever the phase
  transition of a new system is studied. In particular, some aspects of information such as the critical point, the Hamiltonian, and the real
  states of the system are not required for training the 1D NN.
}
\item{The needed configurations considered here for the NN prediction are based on both the spin states and
  the bulk quantities. The storage needed for the bulk observables takes little space.
  Moreover, since the required configurations for the NN prediction
  should be 1D
  lattices made up of 200 sites, in the Monte Carlo simulations
  one can just save randomly 200 spins or several sets of 200 spins.
  Such a procedure takes very small amount of storage space as well.}
  \item{To provide some benchmarks of the performance of the used 1D NN,
  we have conducted several calculations using the standard supervised NN
  procedures. The outcomes indicate potentially at least a factor of several hundred to a few thousand in computational efficiency is gained
  for the 1D NN.
  }
  \item{Due to the high efficiency in both the computation and the storage of the used 1D NN,
    calculations of large system sizes can be done without difficulty. Indeed, the largest linear system size $L$
    considered here for the 2D classical XY model is $L=512$, and all the NN calculations presented in this study
    are conducted on only one server.
    With more data of large system sizes included in the analysis,
    the determined critical points have small uncertainties. For instance, the obtained $\beta_{\text{KT}}$ here for
    the 2D classical $XY$ model is 1.112(6), see also the results calculated in Refs.~\cite{Bea18,Rod19}.
    It should be emphasized the fact that there are many parameters, for instance, the number and the explicit form of filters
    for the convolutional layer, that one can tune to construct a NN. Even NNs with the same infrastructure but initialized differently
    may lead to slightly variant results. Therefore there are some systematic uncertainties not taken into account for any quoted errors
    of NN results, and one should treat the NN outcomes (even those shown here) with caution.  
  }
\item{For the studied 2D generalized classical $XY$ model, the employed NN can efficiently detect the known two phase transitions within
  this model by using only 200 out of 128$^2$ spins.}

\item{Here we demonstrate that with the techniques of NN, quantities other than the helicity
  modulus can be employed to study the BKT type phase transitions as well. Such a method provides an
  alternative and efficient approach to determine the critical points associated with the BKT transitions.}

\item{The majority of NNs considered in the literature have focus on the infrastructure of the NNs in order to
  obtain efficient and working NNs. Here
  our study indicates that by engineering the training and the prediction stages, such as using artificial configurations for the training and
  $\theta\,\text{mod}\,\pi$ instead of $\theta$ for the prediction,
  dramatic improvements are gained. Particularly, a universal NN that is applicable to both the symmetry breaking and the topology related phase
  transitions can be constructed. }
\item{As already being pointed out in Ref.~\cite{Tan21}, the method proposed here is suited for calculating the phase transitions of
  real experimental data. This is because the only required criterion to use our method is the saturation of certain quantities
in both ends of the parameter regions. Such a criterion clearly can be obtained for experiments concerning the phase transitions.}
\end{enumerate}

Moreover, we would like to emphasize the following point.
On the one hand, proposing NN as a new alternative approach for studying various
phases of
matters is extremely impressive and remarkable, since NN traditionally belongs to
completely different fields of science other than the condensed matter physics.
On the other hand,
it will be disappointing if the practical use of NN in investigating critical phenomena
is hindered by some facts such as the requirement of huge amount of computing resources
for the training, a lot of storage space needed for the prediction,
and the necessity of redesignation and retraining whenever a new system
is considered. Here we show that a universal NN trained on a 1D lattice
can be applied to study the phase transitions associated with both
symmetry breaking and topology. The benchmarks
shown here demonstrate that the built
NN is highly efficient in computation.

The (only one) 1D NN used here is shown to be valid for calculating the critical points
of the 3D classical $O(3)$ model, the 3D 5-states ferromagnetic Potts model,
a 3D dimerized quantum spin Heisenberg model, the 2D classical $XY$ model,
and the 2D generalized classical $XY$ model.
It definitely will be interesting to apply our NN, or construct a NN (a supervised
one or an unsupervised one)
following a similar elegant and simple idea, to other models 
that have been studied in the literature
using complicated NN procedures.

Finally, we hope that 
our study may trigger similar NN explorations in research fields other
than the condensed matter physics.

\vskip0.5cm

\section*{Acknowledgements}
Partial support from Ministry of Science and Technology of Taiwan is 
acknowledged.


\begin{thebibliography}{1}



\bibitem{Wan16}
  Lei Wang, 
  Phys. Rev. B {\bf 94}, 195105 (2016).
  
\bibitem{Oht16}
Tomoki Ohtsuki and Tomi Ohtsuki, 
J. Phys. Soc. Jpn. 85, 123706 (2016).

\bibitem{Car16}
  Juan~Carrasquilla, Roger~G.~Melko,
  Nature Physics {\bf 13}, 431–434 (2017).

\bibitem{Tro16}
  Giuseppe Carleo, Matthias Troyer,
  Science 355, 602 (2017).





\bibitem{Chn16}
Kelvin Ch'ng, Juan Carrasquilla, Roger G. Melko, and Ehsan Khatami,
Phys. Rev. X {\bf 7}, 031038 (2017).



\bibitem{Tan16}
  Akinori Tanaka, Akio Tomiya,
  J. Phys. Soc. Jpn. 86, 063001 (2017).

\bibitem{Nie16}
Evert P.L. van Nieuwenburg, Ye-Hua Liu, Sebastian D. Huber,
Nature Physics {\bf 13}, 435–439 (2017).




\bibitem{Den17}
Dong-Ling Deng, Xiaopeng Li, and S. Das Sarma,
Phys. Rev. B {\bf 96} 195145 (2017).



\bibitem{Zha17}
Yi Zhang, Roger G. Melko, and Eun-Ah Kim
Phys. Rev. B {\bf 96}, 245119 (2017). 
 



\bibitem{Hu17}
Wenjian Hu, Rajiv R. P. Singh, and Richard T. Scalettar, 
Phys. Rev. E {\bf 95}, 062122 (2017).

\bibitem{Li18}
  C.-D. Li, D.-R. Tan, and F.-J. Jiang,
  Annals of Physics, 391 (2018) 312-331.




\bibitem{Zha18}
  Pengfei Zhang, Huitao Shen, and Hui Zhai,
Phys. Rev. Lett. {\bf 120}, 066401 (2018).

\bibitem{Bea18}
  Matthew J. S. Beach, Anna Golubeva, and Roger G. Melko,
  Phys. Rev. B {\bf 97}, 045207 (2018).

\bibitem{Lia19}
  Wenqian Lian {\it et al.}
  Phys. Rev. Lett. {\bf 122}, 210503 (2019). 







\bibitem{Rod19}
  Joaquin F. Rodriguez-Nieva and Mathias S. Scheurer,
  Nat. Phys. 15, 790–795 (2019).
  
\bibitem{Car19}
  Giuseppe Carleo, Ignacio Cirac, Kyle Cranmer, Laurent Daudet, Maria Schuld, Naftali Tishby,
  Leslie Vogt-Maranto, and Lenka Zdeborov\'a,
  Rev. Mod. Phys. {\bf 91}, 045002 (2019).


\bibitem{Zha19}
Wanzhou Zhang, Jiayu Liu, and Tzu-Chieh Wei,
Phys. Rev. E {\bf 99}, 032142 (2019).



\bibitem{Don19}
Xiao-Yu Dong, Frank Pollmann, and Xue-Feng Zhang,
Phys. Rev. B {\bf 99}, 121104(R) (2019).


\bibitem{Tan20.1}
  D.-R. Tan { \it et al.}
  2020 New J. Phys. 22 063016. 

\bibitem{Tan20.2}
  D.-R. Tan and F.-J. Jiang,
  Phys. Rev. B {\bf 102}, 224434 (2020).
  
\bibitem{Sin20}
  Japneet Singh, Vipul Arora, Vinay Gupta, and Mathias S. Scheurer,
  arXiv:2006.11868.

  \bibitem{Sch20}
Mathias S. Scheurer and Robert-Jan Slager,
Phys. Rev. Lett. 124, 226401 (2020).



\bibitem{Carrasquilla:2020mas}
J.~Carrasquilla,
Adv. Phys. X \textbf{5}, no.1, 1797528 (2020).




\bibitem{Tomita:2020ylz}
Y.~Tomita, K.~Shiina, Y.~Okabe and H.~K.~Lee,
Phys. Rev. E \textbf{102}, no.2, 021302 (2020).

\bibitem{Baldi:2014pta}
P.~Baldi, P.~Sadowski and D.~Whiteson,
Phys. Rev. Lett. \textbf{114}, 111801 (2015).




\bibitem{Hoyle:2015yha}
B.~Hoyle, 
Astron. Comput. \textbf{16}, 34-40 (2016).


\bibitem{Mott:2017xdb}
A.~Mott, J.~Job, J.~R.~Vlimant, D.~Lidar and M.~Spiropulu,
Nature \textbf{550}, no.7676, 375-379 (2017).

\bibitem{Pang:2016vdc}
L.~G.~Pang, K.~Zhou, N.~Su, H.~Petersen, H.~Stöcker and X.~N.~Wang,
Nature Commun. \textbf{9}, no.1, 210 (2018).

\bibitem{Sha18}
Phiala E. Shanahan, Daniel Trewartha, and William Detmold,
Phys. Rev. D {\bf 97}, 094506 (2018).






\bibitem{Cavaglia:2018xjq}
M.~Cavaglia, K.~Staats and T.~Gill,
Commun. Comput. Phys. \textbf{25}, no.4, 963-987 (2019).

\bibitem{Larkoski:2017jix}
A.~J.~Larkoski, I.~Moult and B.~Nachman,
Phys. Rept. \textbf{841}, 1-63 (2020).




\bibitem{Amacker:2020bmn}
J.~Amacker, W.~Balunas, L.~Beresford, D.~Bortoletto, J.~Frost, C.~Issever, J.~Liu, J.~McKee, A.~Micheli and S.~Paredes Saenz, \textit{et al.}
JHEP \textbf{12}, 115 (2020).



\bibitem{Aad:2020cws}
G.~Aad \textit{et al.} [ATLAS],
Phys. Rev. Lett. \textbf{125}, no.13, 131801 (2020).





\bibitem{Cabero:2020eik}
M.~Cabero, A.~Mahabal and J.~McIver,
Astrophys. J. Lett. \textbf{904}, no.1, L9 (2020).





\bibitem{Nicoli:2020njz}
K.~A.~Nicoli, C.~J.~Anders, L.~Funcke, T.~Hartung, K.~Jansen, P.~Kessel, S.~Nakajima and P.~Stornati,
Phys. Rev. Lett. \textbf{126}, no.3, 032001 (2021).


\bibitem{kera}
https://keras.io

\bibitem{tens}
https://www.tensorflow.org


\bibitem{Tan21}
 D. -R. Tan, J. -H. Peng, Y. -H. Tseng, F. -J. Jiang, arXiv:2103.10846.  




\bibitem{Bin81}
  K. Binder, 
  Z. Phys. B {\bf 43}, 119 (1981).



\bibitem{Has05}  
  Martin Hasenbusch,
  J.~Phys. A {\bf 38} (2005) 5869-5884.

\bibitem{Can14}
Gabriel A. Canova, Yan Levin, and Hefferson J. Arenzon, Phys. Rev. E {\bf 89}, 012126 (2014).
  
\bibitem{Can16}  
Gabriel A. Canova, Yan Levin, and Hefferson J. Arenzon, Phys. Rev. E {\bf 94}, 032140 (2016).
  



\bibitem{Ale20}
Constantia Alexandrou, Andreas Athenodorou, Charalambos Chrysostomou, Srijit Paul,
Eur. Phys. J. B (2020) 93: 226.

\bibitem{Wol89}
U. Wolff, 
Phys. Rev. Lett. {\bf 62}, 361 (1989).




\end{thebibliography}
\end{document}